\documentclass[useAMS,usenatbib]{mn2e}
\usepackage{graphicx,color,epsfig,natbibmnfix}
\usepackage[totalwidth=480pt, totalheight=680pt]{geometry}

\newcommand{\be}{\begin{equation}}
\newcommand{\ee}{\end{equation}}

\newcommand{\apj}{ApJ}
\newcommand{\apjs}{ApJS}
\newcommand{\mnras}{MNRAS}
\newcommand{\aap}{A\&A}
\newcommand{\araa}{ARA\&A}
\newcommand{\apjl}{ApJL}

\def\ltsima{$\; \buildrel < \over \sim \;$}
\def\simlt{\lower.5ex\hbox{\ltsima}}
\def\gtsima{$\; \buildrel > \over \sim \;$}
\def\simgt{\lower.5ex\hbox{\gtsima}}

\def\msun{{\,{\rm M}_\odot}}

\newcommand\mearth{{\,{\rm M}_{\oplus}}}
\def\rsun{{\,R_\odot}}
\def\lsun{{\,L_\odot}}
\def\del#1{{}}

\title[Fu Ori outbursts and planet-star mass transfer]{Fu Ori outbursts and
the planet-disc mass exchange}

\author[Sergei Nayakshin and Giuseppe Lodato]{Sergei Nayakshin$^1$ and Giuseppe
  Lodato$^2$ \\ 
$^1$ Department of Physics \& Astronomy,
  University of Leicester, Leicester, LE1 7RH, UK\\
$^2$ Dipartimento di Fisica, Universit\'a Degli Studi di Milano, Via
  Celoria, 16, Milano 20133, Italy\\
}

\begin{document}

\date{Received}

\pagerange{\pageref{firstpage}--\pageref{lastpage}} \pubyear{2011}

\maketitle

\label{firstpage}

\begin{abstract}
It has been recently proposed that giant protoplanets migrating inward
through the disc more rapidly than they contract could be tidally
disrupted when they fill their Roche lobes $\sim0.1$ AU away from
their parent protostars.  Here we consider the process of mass and
angular momentum exchange between the tidally disrupted planet and the
surrounding disc in detail. We find that the planet's adiabatic
mass-radius relation and its ability to open a deep gap in the disc
determine whether the disruption proceeds as a sudden runaway or a
balanced quasi-static process. In the latter case the planet feeds the
inner disc through its Lagrangian L1 point like a secondary star in a
stellar binary system. As the planet loses mass it gains specific
angular momentum and normally migrates in the outward direction until
the gap closes. Numerical experiments show that planet disruption
outbursts are preceded by long ``quiescent'' periods during which the
disc inward of the planet is empty. The hole in the disc is created
when the planet opens a deep gap, letting the inner disc to drain onto
the star while keeping the outer one stalled behind the planet. We
find that the mass-losing planet embedded in a realistic
protoplanetary disc spawns an extremely rich set of variability
patterns. In a subset of parameter space, there is a limit cycle
behaviour caused by non-linear interaction between the planet mass loss
and the disc hydrogen ionisation instability.  We suggest that tidal
disruptions of young massive planets near their stars may be
responsible for the observed variability of young accreting protostars
such as FU Ori, EXor and T Tauri stars in general.
\end{abstract}


\section{Introduction}\label{sect:intro}


Young protostars can be highly variable \citep[e.g.,][]{Herbig89}. The best
known members of this club are the Fu Orionis objects, which are young stars
showing very large (factors of hundreds) and abrupt increases in their
brightness \citep{HK96}. Observations of these objects are best understood as
variability in the accretion rate onto the growing protostar from a base
accretion rate state of $\sim 10^{-7} \msun$~yr$^{-1}$ to a high of up to
$10^{-4}\msun$~yr$^{-1}$. Although detailed statistics is still lacking, the
duration of the outbursts is believed to be in the range of tens to hundreds
of years. Clear spectroscopic signatures implicate an accretion disc in the
inner $\simlt 1$ AU as the driver of this variable accretion rate and
luminosity output \citep{ZhuEtal07,EisnerH11}. The rise times of the outbursts
are very short, from a year to $\sim$ ten years.  FU Ori phenomenon is not
just a fine detail of young stars evolution: the statistics of FU Ori objects
suggests that an average proto-star may experience 10-20 of such outbursts,
which may still be considered a lower limit on the importance of the FU Ori
outbursts \citep{HK96}. It is not impossible that proto-stars accrete {\em
  most} of its mass during these relatively short but very intense growth
episodes.

A rather natural model for FU Ori variables is a viscously and
thermally unstable accretion disc that undergoes periodic switching
between the stable quiescent and the outburst states
\citep[e.g.,][]{Bell94}. Hydrogen is mainly neutral in the quiescent
state and is mainly ionised in the outburst state, partially
explaining the large difference in the accretion rates. In detail,
however, external triggering of the outbursts may still be desirable
\citep[e.g.,][]{BB92,BellEtal95,LodatoClarke04}, as the theoretically
produced outburst rise time scales are longer than $\sim$ 10 yrs.

The key deficiency of the pure thermal disc instability model
\citep[e.g.,][]{Bell94} is in the fact that the unstable inner disc region is
relatively small, e.g., $R\sim 20\rsun \sim 0.1$ AU \citep{ZhuEtal07}. This
produces outbursts that are too short, unless one uses very small values for
the viscosity $\alpha$-parameter, e.g., as small as $\alpha = 10^{-4}$
\citep{LodatoClarke04}. This seems to be unlikely
\citep{KingEtal07,ZhuEtal09}. In addition, spectral modeling of the inner disc
of the FU Ori variables \citep{ZhuEtal07}, and their NIR interferometry
\cite{EisnerH11}, require the hot disc region to be greater than $R\sim 0.5-1$
AU, which is clearly too large for the pure thermal disc instability models
\citep{ZhuEtal09}. The small extent of the unstable region translates
  into too small a disc mass participating in the instability \citep[since the
    mass is set by the critical surface densities at which the instability
    operates;][]{Bell94}. The disc instability model thus may be said to
  suffer from the ``mass and time-scale deficit problem'' as there is not
  enough mass in the disc to explain the observed long and bright FU Ori
  outbursts.

A very promising way forward to enlarging the unstable region, and thus
solving the deficit problem, was proposed by \cite{ArmitageEtal01}. These
authors pointed out that the intermediate disc region, $R\sim 0.1 - 2$ AU is
expected to have a layered structure \citep{Gammie96}. The disc midplane
region is shielded by the upper layers from the ionisation sources, and is
likely to be too cold to support the Magneto-Rotational Instability
\citep[e.g.,][]{BH98} that would otherwise yield an efficient angular momentum
transport. Accretion thus proceeds only inside the upper ionised ``skin'' of
the disc. Bulk of the material is underneath the skin, too cold to flow
  radially, settling into what is called ``the dead zone''. Over time, and in
  the right range of external mass supply rates, the dead zone becomes very
  massive, e.g., $\sim 0.1 \msun$, which eventually triggers a gravitational
  instability. The associated gravito-turbulence \citep{Lin87,Rice05}
  initiates a rapid angular momentum transfer. \cite{ArmitageEtal01}
obtained relatively bright, $\dot M \sim 10^{-5}\msun$ yr$^{-1}$, outbursts
lasting for as long as $10^3-10^4$ yrs. The current consensus is that the
model of \cite{ArmitageEtal01} satisfies observational and physical
constraints on FU Ori outbursts better than any other
\citep[e.g.,][]{ZhuEtal09}.

Here we propose another solution to the inner disc mass deficit problem, which
relies on an entirely different mass reservoir. We suggest that the material
accreted onto the star during the FU Ori outbursts comes from giant young
planets that fill their Roche lobes inside the inner 1 AU from the star. The
planets are assumed to be pushed inward closer to the star by gravitational
torques of the protoplanetary disc. The significance of young planets as
serious contenders on the role of mass donors for FU Ori outbursts is easily
appreciated. The mass of gaseous proto-planets may be as high as tens of
Jupiter masses ($M_J$), much higher than the likely disc mass in the inner
fractions of an AU. If the material of such a planet is dumped into the inner
disc and then accreted onto the star at an accretion rate of $\sim 10^{-4}
\msun$ yr$^{-1}$, then there is enough gas for a $\sim O(100)$ yrs-long
accretion outburst.

Our work is based on ideas of \cite{Nayakshin11b} who pointed out that the
early ($t\simlt 1$ Myr) stages of the thermal relaxation of the planet are
highly uncertain and model dependent (see a fuller discussion of this in \S
\ref{sec:embryo_contraction}). The ``low density start'' models of such
planets are sufficiently fluffy to be tidally disrupted inside a fraction of
an AU. We shall also find important connections to the earlier work of
\cite{LodatoClarke04} who showed that a massive planet in the inner disc
region can trigger rapid rise outbursts, as observed. Finally, our
  calculations of young protoplanet disruptions are complimentary to the
  ``outburst accretion mode'' simulations of large ($10-1000$ AU) discs by
  \cite{VB05,VB06}. These authors have found that massive gaseous clumps born
  in their discs at $\sim O(100)$ AU distances migrate inward quickly, being
  accreted by their inner boundary condition. 

The purpose of our paper is to investigate the process of tidal disruption of
compact proto-planets in the inner $\sim 0.1$ AU of the star in unison with a
more self-consistent modeling of the protostellar disc evolution to probe the
proposed link of young protoplanets to the FU Ori outbursts.  To model the
disc, we use the standard \citep{Shakura73,Frank02} 1D time-dependent viscous
disc evolution equations in the presence of a massive embedded satellite. An
approximate but well understood and tested type II migration model is used to
communicate the gravitational torques between the planet and the disc
\citep{LinPap86,LodatoClarke04}.  In our modeling of the mass loss, we are
aided by the well known results from semi-detached stellar binaries
\citep[e.g.,][]{Ritter88}.

The structure of the paper and main results are the following. In \S
\ref{sec:analytical} we build a simplified analytical theory for the
mass loss and radial motion of the planet embedded in a
proto-planetary disc. We show that due to conservation of angular
momentum, a mass-transferring planet may migrate outward. Repeating the
well known stellar binary results, we also find that the mass transfer
rate and the migration pattern can be a quasi-steady or run-away
process that may end in a complete (or nearly so) destruction of the
planet. We introduce our numerical methods in \S
\ref{sec:numerics}. In \S \ref{sec:CD} we present numerical
experiments in which the outer disc is replaced by a constant external
torque parameter. This allows us to test our simplified theory in a
``clean'' controlled setting. In \S \ref{sec:full}, simulations with a
self-consistent disc evolution and torque on the planet are
presented. A range of variability patterns is found there, and further
expanded on in \S \ref{sec:lc04}. The effective temperature profiles
of our disc models are discussed in \S \ref{sec:size}. In \S 8 we
summarise the main results of our paper.

\section{Analytical considerations}\label{sec:analytical}

The numerical scheme presented in \S \ref{sec:CD} is well suited for
  studying tidal disruptions of planets in arbitrarily initialised
  time-dependent gas discs. Here we attempt to build a simple analytical
understanding of what to expect from numerical simulations later on. We start
with the simplest possible situations and gradually add different physical
effects.  In this section, the outer accretion disc is only considered as
  a source of a given gravitational torque on the planet ($\tau_e$, cf. \S
  2.3) affecting its radial motion. The inner disc interaction with the planet
  is described by the parameter $f_a$ introduced shortly below. We emphasise
  that $\tau_e$ and $f_a$ are only defined for convenience of our analytical
  study (this section); these parameters are replaced by a self-consistent
  treatment of the disc-planet torques in the time-dependent numerical
  calculation.

\subsection{Outward radial migration driven by planet's mass loss}\label{sec:binary}

A planet embedded in an accretion disc is a subject to gravitational torques
from the surrounding discs that in general push it inward
\citep{LinPap79,GoldreichTremaine80}. Here we point out that a mass-losing
planet can push itself outward. Operationally, the effect is due to a
different balance between the inner and the outer disc torques than that in
the standard no-mass-loss case; however, the effect is best understood simply
on the basis of angular momentum conservation in the star-planet system.

Consider a planet of mass $M_p$ orbiting the star of mass $M_*\gg M_p$ at a
distance $a$ in a circular orbit. The planet's orbital angular momentum is
given by
\begin{equation}
J_p = M_p \Omega_a a^2 = M_p \left(GM_* a\right)^{1/2}\;.
\label{jp}
\end{equation}
If the planet fills its Roche lobe and transfers
mass inward through the Lagrangian L1 point \citep{Frank02}, the system is
analogous to a semi-detached stellar binary system with an extreme mass ratio.
In the case of conservative mass transfer, we have $dM_*/dt = - dM_p/dt >
0$. We neglect the spin angular momentum of the planet as it is small compared
to the orbital angular momentum. We first assume that the star's spin does not
change as a result of gas accretion onto it, in which case the planet's
angular momentum is conserved during the mass transfer process. Therefore,
\begin{equation}
{d \ln a\over dt} = - 2 {d\ln M_p \over dt}\;.
\label{dadt00}
\end{equation}
This demonstrates that the planet migrates outward as it loses mass. This
effect is well known in stellar binary evolution where a conservative mass
exchange between the two components leads to widening of the orbit if the
secondary is less massive than the primary, and shrinking of the orbit if mass
is lost by the more massive star \citep{Ritter88}.

In a more general case, accretion of mass onto the star may spin up
the star. Alternatively, the star could be slowing down its rotation
by passing angular momentum to the inner disc via magnetospheric
torques \citep{CameronCampbell93,ArmitageClarke96}. Also, as we shall
see later, the angular momentum of the gas accreting onto the star may
flow past the planet's orbit if the {\em gap}, which is generally
opened in the disc by a massive planet, is closed. If the viscous time
of the inner disc is long compared with the planet's mass loss
timescale, then we should also account in equation \ref{dadt00} for
the angular momentum stored in the disc. To continue with our
analytical arguments below, we find it convenient to introduce a free
parameter describing the efficiency of the outward migration due to
mass transfer, $f_a>0$:
\begin{equation}
{d \ln a\over dt} = - 2 f_a {d\ln M_p \over dt}\;.
\label{dadt0}
\end{equation}
If the star is a sink of angular momentum, $f_a< 1$, but if it gives the
angular momentum to the disc then $f_a$ may be larger than unity.

We emphasise that, while introduced for convenience in the analytical
considerations described here, $f_a$ is not a free parameter of our
numerical simulations (see Section 3 below), since the outward torque
on the planet due to the inner disc is calculated
self-consistently. We only use $f_a$ in our analytical model below,
and also for an interpretation of the numerical results in cases where
we can make a good guess on the appropriate value of $f_a$.

\subsection{Planet's mass loss rate}\label{sec:loss}

Mass transfer from a Roche lobe-filling secondary to the primary has been a
subject of numerous papers in the context of cataclysmic binaries secular
evolution \citep[e.g.,][]{Ritter88}. The main result of these studies is that
the mass loss rate is a very strong function of $\Delta r = r_p - r_H$, where
\begin{equation}
r_H = a (M_p/3M_*)^{1/3}
\label{rh}
\end{equation}
is the Hills radius of the planet, which we set approximately equal to its
Roche lobe radius, and $r_p$ is the planet's radius. The mass loss rate is
very small for negative $\Delta r$, increasing rapidly for positive $\Delta
r$. 

These basic facts are in fact sufficient for the analytical treatment below,
but we find it timely to introduce our numerical mass transfer scheme here as
well, to emphasise the importance of the difference $\Delta r$ once more.
Before the Roche lobe of the secondary is filled, $\Delta r < 0$, the mass
transfer rate has an exponential form reflecting the exponential decrease of
density with height in the stellar atmosphere \citep[cf. eq. 9
  of][]{Ritter88}:
\begin{equation}
{d M_p\over dt} = - {\rho_{\rm ph} c_{\rm ph} r_H h_p \over e^{1/2}} \exp\left[ 
  {\Delta r \over h_p} \right]\;,
\label{dotM1}
\end{equation}
 where $h_p$ is the scale height of the planet's atmosphere, $\rho_{\rm ph}$
 and $c_{\rm ph}$ are the photosphere's density and sound speed,
 respectively.

When the Roche lobe of the planet is filled, and $\Delta r > 0$, the mass
transfer rate can be calculated in a similar way \citep[see section 1
  of][]{PringleWade85}, and the result turns out to be dependent on whether
the outer layers of the star are convective or radiative
\citep[][]{RappaportEtal83,KR92}. For an isothermal atmosphere, $dM_p/dt
\propto \exp[\Delta r/h_p]$, with $h_p= $const, whereas for a convective star
with polytropic index $\gamma$, the result is
\begin{equation}
{d M_p\over dt} \propto - \Delta r^{(3\gamma-1)/(2(\gamma-1))}\;\quad\hbox{if}\quad\Delta
  r > 0\;
\label{dotM2}
\end{equation}
\citep[equation (1.27) in][]{PringleWade85}. Importantly, for any value of
$\gamma> 1$ this function is strongly increasing with $\Delta r$. For
$\gamma = 5/3$, for example, $dM_p/dt \propto - \Delta r^3$, whereas for
molecular hydrogen, $\gamma \approx 1.4$, and $dM_p/dt \propto - \Delta
r^4$. For definitiveness, we parametrise the mass flow rate through the L1
point as
\begin{equation}
{d M_p\over dt} = - \left({\Delta r\over r_p}\right)^3 {M_p \over t_{\rm
    dyn}}\;\quad\hbox{if}\quad\Delta r > 0\;,
\label{dotM3}
\end{equation}
where $t_{\rm dyn}= \sqrt{r_p^3/GM_p}$ is the (internal) dynamical time of the
planet. Note that when the planet fills its Roche lobe, $t_{\rm dyn} \approx
1/\Omega_k(a) = \sqrt{GM_*/a^3}$, and we can re-write equation \ref{dotM3} as
\begin{equation}
{d M_p\over dt} = - 2 \msun\;\mbox{yr}^{-1} a_{0.1}^{-3/2}\;{M_p \over 10 M_J}
\;\left({\Delta r\over r_p}\right)^3 \;,
\label{dotM4}
\end{equation}
where $a_{0.1} = a/$(0.1 AU). Since the headline factor in equation
\ref{dotM4} is so large, one obtains very high Roche lobe overflow rates even
for modest values of $\Delta r$. For example, even at $\Delta r = 0.01 r_p$
the overflow rate is about two order of magnitude larger than the mean
accretion rate onto embedded protostars, $\dot M \sim 10^{-8} \msun$~yr$^{-1}$
\citep{HartmannEtal98}.

We found through experimenting with different power-law indexes in
this dependence that the results are very insensitive to the exact
form of the mass loss rate dependence on $\Delta r$. This result is
well known from stellar binary evolutionary calculations, and will be
re-derived for the problem at hand below. Physically, since the mass
transfer rate is a strong function of $\Delta r$, it turns out that
the process is stable only if $|\Delta r|$ may be maintained very
small, which is equivalent to requiring $r_p \approx r_H$. In this
situation the mass loss rate self-adjusts to satisfy the angular
momentum loss or gain constraints \citep[cf.][]{Ritter88}. Different
mass loss formulations only yield slightly different values of $\Delta
r$ at which the required equilibrium value of $dM_p/dt$ is achieved.

On the other hand, in cases where the mass transfer leads to $r_H$ shrinking
faster than $r_p$, or $r_p$ increasing faster than $r_H$, a runaway mass
transfer occurs. One well known example of this in stellar binaries is the
case where the secondary is more massive than the primary, in which case the
separation of the components shrinks rather than increases as the mass is
transferred \citep{PringleWade85}. In this case the disruption of the
companion is nearly dynamical, and therefore the exact form of the transfer
rate is again not very important.

\del{We shall now investigate the conditions under which the mass transfer process
is quasi-stable, following the treatment presented in \S 2.2 of
\cite{Ritter88}.}

\subsection{Roche lobe overflow: L1 or both L1 and L2?}\label{sec:l1}

We follow here the usual assumption that the planet's surface is
determined by the equipotential surfaces \citep[e.g., \S of][]{Frank02}. We
also assume that the planet and the star are on circular orbits around the
centre of mass of the system. The dimensionless potential in a frame
corotating with the centre of mass of the star-planet system has two saddle
points, L1 and L2. The points lie on the line connecting the star and the
planet, with L1 positioned between the planet and the star, and L2 further
away ``behind'' the planet.

In stellar binaries, the overflow of the secondary's Roche lobe always
proceeds via the Lagrangian point L1 since it is much closer to the
secondary. However, for the planet-star system, the mass ratio, $q = M_p/M_*$, is
very small, and thus the overflow may proceed via both L1 and L2. We quantify
this following the treatment given in \S 4.1 of \cite{GuEtal03}. The
difference between distances
of the L1 and L2 points ($D_1$ and $D_2$, respectively) from the centre of the
planet are
\begin{equation}
\Delta D \equiv D_2 - D_1 \approx {2 \eta a\over 3} \left(\eta^{1/2} + \eta\right) \;,
\end{equation}
where $\eta = r_H/a = (M_p/3M_*)^{1/3}\ll 1$ \citep[see eq. 63
  in][]{GuEtal03}.

We see that the point L2 is distance $\Delta D$ further away from the centre
of the planet than L1. Now, if the atmosphere's scaleheight $h_p$ were
infinitely small, the overflow would only proceed via the L1 point (if the
planet overflow maintains the $r_p = r_H \approx D_1$ condition). However, if
the atmosphere is extended enough, that is $h_p \simgt \Delta D$, then the
outflow should proceed through both points L1 and L2.

Equation (66) of \cite{GuEtal03} shows that
\begin{equation}
{h_p \over \Delta D} \approx {9\over 8} \;{c_{\rm ph}^2 a \over GM_p}\;.
\label{h_over_d}
\end{equation}
Our models will characteristically assume that $T_{\rm eff} \sim 10^3$ K (see
\S \ref{sec:embryo_contraction} below), and the planet-star separation at
which the Roche lobe overflow occurs is often $a\sim 0.1$ AU. For these
fiducial parameters,
\begin{equation}
{h_p \over \Delta D} \approx 0.45 \;\left({T_{\rm eff}\over 10^3\;\mbox{K}}\right)
\;\left({a\over 0.1 \mbox{AU}}\right)\; \left({M_J \over M_p}\right)\;.
\label{h_over_d_fid}
\end{equation}
This shows that for massive young protoplanets, $M_p \sim 10 M_J$, the
overflow should indeed proceed mainly via the L1 point, but as the planet's
mass drops below $1M_J$ both L1 and L2 can be letting the planet's atmosphere
to escape.

There is actually one further condition to check. In \S \ref{sec:loss} we have
shown that the overflow rate is a strongly increasing function of $\Delta
r$. In the simulations below, we shall find relatively large Roche lobe
overflow rates, requiring small but not infinitesimally small values for
$\Delta r/r_p$. The difference between locations $L1$ and $L2$, $\Delta D$, is
also small in units of $r_p$. For a large enough values of $|dM_p/dt|$, it is
possible that the required Roche lobe ``overfilling'', $\Delta r$, is larger
than $\Delta D$. We expect that if $\Delta r \simlt \Delta D$, the overflow
proceeds mainly via L1 as argued above, but if $\Delta r \gg \Delta D$, both
L1 and L2 will siphon gas away at comparable rates. To compare these, we note
that for $M_p = 10 M_J$, $\Delta D\approx r_p \eta^{3/2} = 0.06 r_p (M_p/10
M_J)^{1/2}$. On the other hand, the planet mass loss of $|dM_p/dt| = 10^{-4}
\msun$~yr$^{-1}$ requires $\Delta r \approx 0.037 r_p$. Thus we see that, at
least for $M_p = 10 M_J$ the overflow should proceed mainly through the L1
point.

In general, the ratio of the two length scales is
\begin{equation}
{\Delta r \over \Delta D} \approx 0.6 \; \left[{|dM_p/dt| \over 10^{-4}
\msun \mbox{yr}^{-1}}\right]^{1/3} \; a_{0.1}^{1/2} \left({10 M_J \over M_p}\right)^{-5/6}\;.
\label{r_over_d}
\end{equation}
This shows that at the high mass loss rates the outflow proceeds via L1 point
as long as $M_p \simgt 5 M_J$; at lower planet masses the outflow may occur
via both L1 and L2 points. 

We conclude that in the most typical parameter space sampled by our models
below, the Roche lobe overflow occurs via the L1 point. However, for
relatively low mass planets, $M_p \simlt 1 M_J$, the overflow should take
place via both L1 and L2. In this paper we make an approximation that the
outflow always occurs via L1.

\subsection{Definitions}\label{sec:define}

The time evolution of $\Delta r$ is of primary interest to us
here. Let us calculate the full time derivatives of the Hill (Roche) radius and the
planet radius:
\begin{equation}
{d \ln r_H \over dt} = \left({\partial \ln r_H \over \partial \ln M_p} \right) {d
  \ln M_p \over dt} + \left({\partial \ln r_H \over \partial t} \right)_{\dot
  M_p=0}\;, 
\label{drh_dt}
\end{equation}
and 
\begin{equation}
{d \ln r_p \over dt} = \left({\partial \ln r_p \over \partial \ln M_p} \right)
{d \ln M_p \over dt} + \left({\partial \ln r_p \over \partial t} \right)_{\dot
  M_p=0}\;,
\label{drp_dt}
\end{equation}
respectively. Here, $(\partial \ln r_p/ \partial \ln M_p)$ is the adiabatic
mass-radius exponent of the planet, discussed below, and $(\partial\ln r_p/
\partial t)_{\dot M_p=0}$ is the time derivative of planet's radius due to
thermal relaxation, e.g., radiative cooling of the planet. Similarly,
$(\partial \ln r_H/ \partial \ln M_p)$ is the rate of Hills radius change due
to planet's mass changing, and $(\partial\ln r_H/ \partial t)_{\dot M_p=0}$ is
the term due to external torques onto the planet in the absence of the mass
loss.

Since $r_H = a(M_p/3M_*)^{1/3}$, and in view of equation \ref{dadt0},
\begin{equation}
\zeta_H \equiv {\partial \ln r_H \over \partial \ln M_p} \approx -2f_a +
1/3\;,
\label{zetah}
\end{equation}
where we neglected the small term $-(\partial \ln M_*/\partial \ln M_p) = q =
M_p/M_* \ll 1$. Further, we introduce the following notations for brevity
\begin{equation}
\zeta_p \equiv {\partial \ln r_p \over \partial \ln M_p}\;,
\label{zetap}
\end{equation}
\begin{equation}
\tau_c = \left({\partial \ln r_p \over \partial t} \right)_{\dot
  M_p=0}^{-1}\;, 
\label{tauc}
\end{equation}
\begin{equation}
\tau_{\rm e} = \left({\partial \ln r_H \over \partial t} \right)_{\dot
  M_p=0}^{-1} = \left({\partial \ln a\over \partial t} \right)_{\dot
  M_p=0}^{-1}\;. 
\label{taue}
\end{equation}

The mass-radius relation for a polytropic planet is given by
\begin{equation}
r_p \propto M_p^\frac{1-n}{3-n} \;,
\label{mass_radius}
\end{equation}
where $n=1 /(\gamma-1)$ is the polytrope's index, and $\gamma$ is the ratio of
the specific heats for the planet's gas. It is noteworthy that the radius-mass
relation for a polytrope strongly depends on $\gamma$, and that the polytropic
cloud usually (for large enough values of $\gamma$) expands as the mass is
lost. For example, for $\gamma = 5/3$, $n= 3/2$, and $r_p\propto M_p^{-1/3}$;
for $\gamma = 7/5$, $n = 5/2$, $r_p\propto M_p^{-3}$.

Note that $\tau_{\rm e}$ can be related to the external torque, $T_{\rm ext}$,
on the planet. Here we anticipate that in our model situation the external
torque is negative, $T_{\rm ext}<0$, pushing the planet closer to the star. To
relate $\tau_{\rm e}$ and $T_{\rm ext}$, note that angular momentum
conservation in the absence of mass loss yields
\begin{equation}
M_p {d \over dt} \left(GM_* a\right)^{1/2} = -|T_{\rm ext}|\;,
\end{equation}
where $(GM_*a)^{1/2} = \Omega_a a^2$ is a specific angular momentum of the
planet. From which we conclude that
\begin{equation}
\tau_{\rm e}^{-1} = - {2 |T_{\rm ext}|\over M_p \Omega_a a^2}<0
\label{tau_ext}
\end{equation}

Note also that $\tau_{\rm c}$ is negative for a planet contracting as the
result of radiative cooling and it is positive for a planet expanding with
time at a constant mass. The latter situation may occur when the planet's
interior is heated by the solid core accretion luminosity
\citep{PollackEtal96,Nayakshin10b}, or due to dissipation of tides induced by
the parent star \citep[e.g.,][]{GuEtal03}, irradiation
\citep{BurrowsEtal00,BaraffeEtal03}.

\subsection{Stability of mass transfer}\label{sec:simplest}

In accord with our plan to progress from simple to more complex we first study
a non-cooling planet without external torques, i.e., $1/\tau_{\rm e} =
1/\tau_c = 0$, which has $R_H \approx R_p$, i.e., transfers mass onto the star
at some non-zero rate, $-\dot M_p^{(0)}$. As far as realistic planets in
massive gas discs are concerned, this model situation is of an academic
interest, as one needs external torques or internal evolution of the planet
(e.g., planet swelling due to internal or tidal heating) to fill its Roche
lobe, but it will be seen to provide important hints to the results of the
more realistic numerical experiments later on.

Following the discussion in \S \ref{sec:loss}, the dominant factor in the mass
loss rate dependence is the difference $r_p - r_H$. Write
\begin{equation}
{d \ln M_p \over dt} \approx -  \phi\left(\Delta r\right)\;,
\end{equation}
where $\phi>0$ is a positive and monotonically increasing function of
its argument (note that $dM_p/dt < 0$). We have explicitly neglected
here the much slower dependence of the mass transfer rate on other
variables (such as the photosphere's temperature and the density in
equation \ref{dotM1}). Within this approximation, we have,
\begin{equation}
{d^2 \ln M_p\over dt^2} = - \phi'\left ({d r_p\over
  dt} - {d r_H\over dt}\right) \;,
\label{dmdt21}
\end{equation}
where $\phi' = \partial \phi/\partial \Delta r$.
Since at the onset of the mass transfer $r_p \approx r_H$,
\begin{equation}
{d^2\ln M_p \over dt^2}  = - \phi' r_p \left ({d \ln
  r_p\over dt} - {d \ln r_H\over dt}\right) \;.
\label{dmdt22}
\end{equation}
Referring now to equations \ref{drh_dt}, \ref{drp_dt}, and definitions
\ref{zetah}, \ref{zetap}, we find that
\begin{equation}
{d^2 \ln M_p\over dt^2}  = - \phi' r_p \left(\zeta_p
  - \zeta_H\right) {d \ln M_p\over dt}\;.
\label{dmdt23}
\end{equation}

It is now apparent that the stability of the mass transfer rate is determined
by the sign of $\zeta_p-\zeta_H$ since $ \partial \phi'> 0$. If this term is
negative, the mass loss rate increases exponentially with time. Thus, the mass
loss is unstable if
\begin{equation}
\zeta_p-\zeta_H = \zeta_p -{1 \over 3} + 2f_a <0\;,
\label{unstable}
\end{equation}
where we used equation \ref{zetah}.  This relation demonstrates that stability
of the mass loss process is decided by the mass-radius relation for the
planet, and by the angular momentum transfer between the planet and the star
(the factor $f_a$). Physically, this is because the planet moves outward as
the result of the mass transfer, which increases $r_H$. However, the planet
may also increase in size due to mass loss.  When the condition
(\ref{unstable}) is satisfied, $r_p$ increases faster than $r_H$ as the planet
siphons away its mass. In this case, once the planet filled its Roche lobe
(Hill's radius), the mass loss process is a runaway one.

The opposite situation, $\zeta_p-\zeta_H > 0$, is stable. In the presently
considered case of zero external torques and no cooling, the mass loss rate
drops exponentially with time. The physics of this is simple -- the planet
moves outward and eventually it becomes smaller than its Roche lobe. In a more
realistic situation where there are non-zero external torques, a steady-state
situation can be set up, where the planet moves radially at just the right
rate to maintain the balance $r_p \approx r_H$.

Clearly, the stability of the mass transfer is strongly dependent both
on the details of the evolution of the planet's angular momentum
(described by the parameter $f_a$) and by the evolution of the
planetary radius in response to mass loss (described by the parameter
$\zeta_p$). 

\del{For most of the cases described here, we will adopt a
simple mass-radius relation for polytropic spheres to determine
$\zeta_p$. For example, for a polytropic sphere with index
$\gamma=5/3$, we have $\zeta_p=-1/3$.}

\subsection{Quasi equilibrium mass loss}\label{sec:quasi_steady}

Let us now turn our attention to more realistic situations where the external
torque and the planet's radius internal evolution are not negligible. In this
case equation \ref{dmdt22} leads to
\begin{equation}
{d^2\over dt^2} \ln M_p = -  \phi' r_p
\left[\left(\zeta_p - \zeta_H\right) {d \ln M_p\over dt} + {1\over \tau_c} -
  {1\over \tau_{\rm e}}\right]\;.
\label{dmdt2g}
\end{equation}
Since this is a linear differential equation, our conclusions about the
stability of mass transfer do not change. If $\zeta_p -\zeta_H < 0$, the mass
transfer is unstable as the mass loss rate runs away exponentially. 

However, if $\zeta_p - \zeta_H > 0$, there is a quasi steady state solution of
the above equation with $d^2 \ln M_p/dt^2\approx 0$, where the mass loss rate
is locked at a unique value given by
\begin{equation}
\dot M_{\rm eq} = - {M_p \over \zeta_p -\zeta_H}\left({1\over \tau_c} -
     {1\over \tau_{\rm e}}\right)\;.
\label{dotm_eq}
\end{equation} 
This result, except for different notations, is the same as equation (16) of
\cite{Ritter88}, a well known result in the binary mass transfer studies.

We note that if the planet is cooling and there is no internal energy source,
then $\tau_c < 0$, e.g., the planet contracts with time (equation
\ref{tauc}). The external torque is also negative by assumption, therefore for
a {\em contracting} planet, 
\begin{equation}
\dot M_{\rm eq} = - {M_p \over \zeta_p -\zeta_H}\left(-{1\over |\tau_c|} +
     {1\over |\tau_{\rm e}|}\right)\;.
\label{dotm_eq_c}
\end{equation} 
As $\dot M_p < 0$ by definition, we see that the quasi-equilibrium mass loss
solution is possible only if $|\tau_c| > |\tau_e|$. In the opposite case the
planet contracts too quickly to be tidally disrupted \citep{Nayakshin11b};
despite migrating in closer to the star, the mass transfer rate shuts down
with time because $r_p$ decreases faster than $r_H$.

On the contrary, if there are evolutionary reasons for the planet to swell
with time, such as stellar irradiation or a powerful internal energy release
by the solid core, then the thermal relaxation term in equation \ref{dotm_eq}
amplifies rather than dumps the mass loss rate. In fact, mass loss from the
planet may occur in this case even if there are no external torques on
the planet.

We can now justify our claim (also well known from the stellar binary mass
transfer studies) in the end of \S \ref{sec:loss} that the equilibrium mass
transfer rate does {\em not} depend on the exact form of the mass transfer
rate function, $\phi$. As long as that dependence is monotonic and a strong
one, the difference $\delta r = r_p-r_H$ adjusts to ``the right value'' which
yields the equilibrium mass transfer rate given by equation \ref{dotm_eq}.
The mass transfer rate in this sense is not a ``driving'' variable of the
problem; it is one determined by the balance of the torques on the planet.

Curiously, in the case of the external torques strongly dominating over the
thermal relaxation of the planet, $|\tau_{\rm e}|\ll |\tau_c|$, the mass
transfer rate is further insensitive even to the planet's mass,
$M_p$. Consulting equation \ref{tau_ext}, we find that equation \ref{dotm_eq}
reads in this case,
\begin{equation}
\dot M_{\rm eq} = - {2 |T_{\rm ext}| \over \zeta_p -\zeta_H} {1\over \Omega_a a^2} 
\label{dotm_ext}\;.
\end{equation}
That is, in this case, the mass loss rate depends only on the external torque
and location of the planet, $a$, but not the mass of the planet.

\subsection{Radial migration due to mass loss}\label{sec:m_e}

Having understood the conditions under which the mass transfer rate can be
steady, we now turn to the consequences of this for the radial migration of
the planet.  Adding the external torque on the planet to equation \ref{dadt0},
we have
\begin{equation}
{d \ln a\over dt} = - 2 f_a {d\ln M_p \over dt} + {1 \over \tau_{\rm e}}\;.
\label{dadt0e}
\end{equation}
When the mass loss rate is in the steady state regime (equation
\ref{dotm_eq}), this becomes
\begin{equation}
{d \ln a\over dt} = {1 \over \tau_{\rm e}} {\zeta_p -1/3\over \zeta_p -
    1/3 + 2f_a} + {1 \over \tau_{\rm c}}{2f_a\over \zeta_p -
    1/3 + 2 f_a}\;.
\label{dadt2e}
\end{equation}
We remind ourselves that $\tau_{\rm e}$ is defined to be negative (equation
\ref{tau_ext}), that is, the the external torques are defined to drive
  the planet in rather than out. In contrast, $\tau_{\rm c}$ could be
positive or negative. The remarkable property of equation \ref{dadt2e} is
that, depending on the different terms in the equation, the planet may migrate
either in or out {\em despite being pushed inward by the external torque,
  e.g., the outer disc}. Furthermore, the relative magnitude of these terms
may vary as the planet loses mass or evolves internally. We shall later see
that due to this, the same planet may change the direction of migration
several times.

\subsubsection{Negligible cooling: inward or outward
  migration?}\label{sec:in_or_out}

When the thermal evolution of the planet is insignificant during the tidal
disruption process, so that $|\tau_c| \gg |\tau_e|$, we can neglect the cooling
term in equation \ref{dadt2e}:
\begin{equation}
{d \ln a\over dt} = {1 \over \tau_{\rm e}} {\zeta_p -1/3\over \zeta_p -
    1/3 + 2f_a} \;.
\label{dadt2c0}
\end{equation}
Recall that in the steady-state mass transfer rate regime, the denominator of
equation \ref{dadt2c0} is positive (cf. \S \ref{sec:simplest}). Since $\tau_e <
0$, we see that when the polytropic approximation for the planet is adequate,
\begin{equation}
\hbox{sign}\left({d a\over dt}\right) = - \hbox{sign}\left(\zeta_p-1/3\right)\;.
\label{in_or_out}
\end{equation}
That is, the planet migrates outward if $\zeta_p < 1/3$, and inward
otherwise. In particular, polytropic planets that expand as they lose mass
($\zeta_p < 0$) always migrate outward. Physically, $\zeta_p$ controls the
amount of mass that the planet has to shed in response to being forced inward
by the external torque. This response is meager for planets that shrink due to
mass loss. In order to maintain the $r_p\approx r_H$ equilibrium
  condition, such planets may simply lose mass as $r_H$ shrinks. Thus
  external torques are able to push these planets in. In the opposite case,
  when $\zeta_p <0$, applying the external torque to push the planet in causes
  it to lose so much mass that the outward torque of this material exceeds
  the originally applied torque, and the planet thus moves outward rather than
  inward.

\subsubsection{Rapidly cooling planets}\label{sec:cooling} 

When thermal relaxation of the planet is not negligible, e.g., $|\tau_c|$
comparable to or much smaller than the external torque time scale, $
|\tau_e|$, the last term in equation \ref{dadt2e} cannot be neglected.  A
casual look at the equation \ref{dadt2e} would suggest that a rapidly
contracting planet, $\tau_c<0$, $|\tau_c| \ll |\tau_e|$, migrates as
prescribed by the last term of that equation. However, as remarked after
equation \ref{dotm_eq_c}, the quasi-equilibrium mass loss is only possible if
$|\tau_c| > |\tau_e|$ or else the rapid thermal contraction of the planet
shuts down the mass transfer entirely because $r_p$ becomes smaller than
$r_H$.

Nevertheless, even in the quasi-equilibrium situation the second term in
equation \ref{dadt2e} may indeed exceed the first one and hence dictate the
direction of the planet's migration.  To give an example, consider our standard case of
a planet with $\zeta_p = -1/3$ (gas polytropic index $\gamma = 5/3$). Equation
\ref{dadt2e} reads for this case
\begin{equation}
{d \ln a\over dt} = - {1 \over \tau_{\rm e}} {1\over 3 f_a -1} + 
{1 \over \tau_{\rm c}} {3 f_a\over 3 f_a -1}\;.
\label{dadt3e}
\end{equation}
Therefore we see that for $|\tau_c|$ somewhat larger than $ |\tau_e|$, the
planet may lose the mass in the quasi-steady regime and yet migrate outward
if $f_a$ is sufficiently close to unity. 

\subsubsection{Thermally expanding planets}\label{Sec:exp}

Let us now turn to the case of planets expanding as the result of thermal
evolution, e.g., an inner energy source such as the solid core formation due
to dust sedimentation \citep{Nayakshin10b}. By definition, $\tau_{\rm c} > 0$
in this case, so the steady-state mass loss is possible for any value of
$\tau_{\rm e}$ (provided $\zeta_p-\zeta_H>0$). The last term in equation
\ref{dadt2e} acts to push the planet outward; however, depending on the first
term in that equation the migration direction can be inward or outward.

\section{Numerical Methods}\label{sec:numerics}

\subsection{Mass and angular momentum transfer in the disc}\label{sec:m_transfer}

We use the code of \cite{LodatoEtal09} for the evolution of an accretion disc
in the presence of an embedded satellite, rescaled to the case of a protostar
of mass $M_*$ and a planet of mass $M_{\rm p}$. The disc is described by a
diffusive evolution model that includes the tidal torque term arising from the
planet and the mass deposition term, related to the planet mass loss:
\begin{eqnarray}
\nonumber
  \frac{\partial\Sigma}{\partial
  t} & = &\frac{3}{R}\frac{\partial}{\partial R}
  \left[R^{1/2}\frac{\partial}{\partial R}(R^{1/2}\nu\Sigma)\right]
  -\frac{1}{R}\frac{\partial}{\partial
  R}\left(2\Omega R^2\lambda\Sigma\right) \\
 & - & \frac{\dot M_p}{2\pi R}\;\delta\left(R-R_{\rm
   dep}\right) \qquad
\label{eq:diffplanet}
\end{eqnarray}
where $\lambda=\Lambda/(\Omega R)^2$, and $\Lambda$ is the specific tidal
torque, where
\begin{eqnarray}
\label{eq:torque}
\lambda= & \displaystyle\frac{q^2}{2}\left(\displaystyle
\frac{a}{p}\right)^4  &
R>a \\
\nonumber \lambda= & -\displaystyle\frac{q^2}{2}\left(\displaystyle 
\frac{R}{p}\right)^4  & R<a. 
\end{eqnarray}
In equation (\ref{eq:torque}), $\Omega = \sqrt{GM_*/R^3}$ is the angular
velocity at radius $R$, $a$ is the radial position of the planet, $q=M_p/M_*$
and $p=R-a$. This simplified form of the specific torque is commonly used in
literature (see, e.g.,
\citealt{LinPap86,armibonnell02,LodatoClarke04,AlexanderEtal06}). We smooth
the torque term for $R\approx a$, where it would have a singularity (see
equation \ref{eq:torque}). We use the same smoothing prescription as in
\citet{SyerClarke95} and \citet{LinPap86}, i.e. for $|R-a|<\max[H,r_H]$, where
$H$ is the disc thickness and $r_H = a(M_{\rm p}/3M_*)^{1/3}$ is the size of
the Hill sphere (Roche lobe) of the planet.

We shall note here that we use a constant $\alpha$-viscosity parameterisation
for the angular momentum transfer in the disc. Gravitoturbulent discs could in
principle generate an additional viscosity through gravitational torques
\citep{Lin87,Gammie01,Rice05}. However, here we are interested in the inner
few AU discs that are unlikely to be self-gravitating since the radiative
cooling is very inefficient in units of local dynamical time in such discs
\citep[e.g.,][]{Rafikov05,ClarkeLodato09}. Such small discs are also expected
to be much less massive than the central star and hence the global model of
mass transfer is also very unlikely \citep{LodatoRice05,CossinsEtal09}.

The last term in equation \ref{eq:diffplanet} describes the mass deposition in
the inner disc by the planet (note that $\dot M_p < 0$). The mass deposition
takes place into a narrow ring centred onto the deposition radius $R_{\rm
  dep}$, which is the circularisation radius of the planet's material lost
through the L1 point into the inner disc. The specific angular momentum of the
gas at L1 is $\Omega_a (a-r_H)^2$, where $\Omega_a = \sqrt{GM_*/a^3}$, and
thus the deposition radius is given by
\begin{equation}
R_{\rm dep} = a \left(1 - \frac{r_H}{a}\right)^4\;.
\label{rdep}
\end{equation}
Operationally, we find the radial zone inside of which $R_{\rm dep}$ fall and
deposit all of the mass lost by the planet there. Note that $R_{\rm dep}$ is a
function of time. Depositing gas in a single zone does not appear to lead to
any numerical problems as the gas spreads in both directions from that
locations fairly quickly, resulting in a well behaved surface density profile.

\subsection{Thermal disc equations}\label{sec:therm_balance}

The rest of disc equations follow the standard vertically-averaged approach
\citep[e.g.,][]{Shakura73} with minimum necessary complexity. In particular,
disc viscosity in the $\alpha$-prescription for a gas--pressure--dominated
disc is
\begin{equation}
\label{eq:viscosity}
\nu = \alpha c_{\rm g}^2/\Omega,
\end{equation}
where $c_{\rm g}=kT_{\rm c}/\mu m_H$ is the gas sound speed at the
disc midplane, $T_{\rm c}$ is the midplane temperature, $\mu$ is the
dimensionless mean molecular weight and $m_H$ is the mass of the
proton. The disc vertical scale height $H$ is found by solving for the
vertical pressure balance, taking into account the radiation pressure
(although for most cases considered here radiation pressure is
generally negligible):
\begin{equation}
P_{\rm gas} + {1\over 3} a_{\rm rad} T_{\rm c}^4  = \frac{GM_*}{R}\rho_c
  \left({H \over R}\right)^2\;,
\label{hydrostatic}
\end{equation}
where $\rho_c $ and $P_{\rm gas} = \rho_c kT_{\rm c}/\mu m_H$ are the gas
midplane density and pressure, respectively, and $a_{\rm rad}$ is the
radiation energy density constant. The central gas density and the disc
surface density are related by $\Sigma = 2 \rho_c H$.

The midplane equilibrium temperature is solved assuming the vertical energy balance
for the disc:
\begin{equation}
\frac{9}{8}\nu(T_{\rm eq})\Sigma\Omega^2 = {\sigma_{B}T_{\rm eq}^4\over
  \tau}\;,
\label{temp_eq}
\end{equation}
where $\tau = \kappa(\rho_c, T_{\rm eq}) \rho_c H$ is the disc opacity for
the opacity coefficient $\kappa(\rho_c, T_{\rm eq})$. For opacities varying
slowly with temperature, this approach is entirely satisfactory, as the
thermal equilibrium is established on the time scale $t_{\rm th} = 1/\alpha
\Omega$, which is much shorter than the disc viscous time \citep[e.g.,
  cf.][]{Frank02}. 

We modify this standard equation to account approximately
for the irradiation heating of the disc by the star, whose luminosity is set
to $L_* = \lsun (M_*/\msun)^3$. The incidence angle of the irradiation for
simplicity is set to give $\cos i = 0.1$ everywhere, although in principle this
should be a function of position and the disc vertical scale height,
$H(R)$. Thus we write 
\begin{equation}
T_{\rm irr} = \left({L_* \cos i\over 4 \pi \sigma_B R^2 }\right)^{1/4} \;,
\label{tirr}
\end{equation}
and
\begin{equation}
T_{\rm eq}^4 = T_{\rm irr}^4 + \frac{9}{8\sigma_B}\tau \nu\Sigma\Omega^2 \;.
\label{temp_eq2}
\end{equation}

For protostellar discs, however, due to a strong temperature dependence of
opacity coefficient $\kappa$ on temperature (primarily), the disc may be
thermally unstable \citep[e.g.][]{Bell94}. In terms of equation
\ref{temp_eq2}, it means that there may be more than one solution for a given
$\Sigma$, and that some of the solutions may be thermally unstable. The time
evolution of the disc in this case depends on its previous thermal state, and
one needs to solve a time-dependent version of the energy equation rather than
assume thermal equilibrium.

While thermal instabilities are clearly relevant to the problem at hand, we
would like to concentrate here on the ``new'' physics -- the planet tidal
disruption and the resulting inner disc refilling -- as much as possible.  We
therefore pick a very simple and transparent numerical method to deal with
thermal instabilities in the present paper. In this minimalist approach,
similar to that of \cite{LodatoClarke04}, we use a time-dependent energy
balance equation with a radial advection of heat term:
\begin{equation}
{\partial T\over \partial t} = - {T - T_{\rm eq} \over t_{\rm th}} - v_R
{\partial T\over \partial R}\;.
\label{T_of_t}
\end{equation}
In this equation, $T_{\rm eq}$ is found from equation \ref{temp_eq2} using the
previous timestep value for the gas temperature, $T_t$, in the disc opacity
calculation, $\kappa(\rho_c, T_t)$. The radial velocity $v_R$ is given by the
following:
\begin{equation}
v_R = - \frac{3}{\Sigma R^{1/2}} {\partial\over \partial R}\left[\nu \Sigma
  R^{1/2}\right] + 2\Omega R \lambda\;.
\label{vr_disc}
\end{equation}
Our simple energy equation allows us to capture the thermal disc
  instability if it does occur without an excessive associated numerical
  cost.

\subsection{Radial motion of the planet}\label{sec:radp}

In our approach, there are two ways in which the angular momentum is passed
between the planet and the disc. First of these is due to gravitational
torques (the second term on the right hand side of equation
\ref{eq:diffplanet}) and is independent of the instantaneous mass loss rate by
the planet. The term represents the back reaction of the disc on the planet's
motion and follows from angular momentum conservation
\citep[see][]{LodatoEtal09}:
\begin{eqnarray}
\label{eq:planet}
M_p \left(\frac{\mbox{d}}{\mbox{d} t}\Omega_{a}a^2\right)_{\rm J} & = &
-\int_{R_{\mathrm{in}}}^{R_{\mathrm{out}}}2\pi \Omega^2 R^3\lambda\Sigma d R
\\ \nonumber   & = &  -2\pi GM_* \int_{R_{\rm in}}^{R_{\rm out}}\lambda\Sigma
dR,
\end{eqnarray}
where the integral is taken over the whole disc surface and $\Omega_a$
is the angular velocity of the planet. The subscript ``J'' indicates
that the term arises due to the planet-disc angular momentum exchange.

The second channel for the angular momentum exchange is only present when the
planet loses mass to the disc. As explained in \S \ref{sec:m_transfer}, the
disc gains material with the specific angular momentum of the L1 point of the
planet. The material is deposited at the respective circularisation radius,
$R_{\rm dep}$ (equation \ref{rdep}). The planet's specific angular momentum 
thus increases at the rate
\begin{equation}
\label{eq:planet2}
M_p \left(\frac{\mbox{d}}{\mbox{d} t}\Omega_{a}a^2\right)_{\rm mass\, loss} =
-{\dot M_p}\left[\Omega_a a^2 - \Omega_a (a-r_H)^2\right]\;.
\end{equation} 

Adding the two terms for the radial migration of the planet, we obtain
\begin{eqnarray}
\frac{\mbox{d}\ln a}{\mbox{d} t} & = &
-  \frac{4\pi G}{\sqrt{GM_*a}} \int_{R_{\rm in}}^{R_{\rm
    out}}\lambda\Sigma dR
\\ \nonumber &  - & \quad \frac{\dot M_p}{M_p}\frac{2r_H}{a}\left(2 - \frac{r_H}{a} \right)\;.
\label{eq:planet_tot}
\end{eqnarray}
Note that the planet mass loss rate appears directly only in the second term
on the right hand side, but it is also indirectly present in the first one. As
the planet fills in the inner disc with gas, it changes the disc surface
density profile, and therefore the first term on the right as well. This
system of equations is thus quite non-linear.

\subsection{Planet contraction}\label{sec:embryo_contraction}

In this paper we are only interested in the planets disrupted inside the inner
1 AU of the star. Such disruptions were termed ``hot'' by \cite{Nayakshin11b}
due to their possible physical links to the ``hot'' jupiters and lower-mass
close-in exoplanets. Only giant planets in which molecular hydrogen has been
dissociated into atomic or ionised H are dense enough to venture into this
region \citep[see][]{Nayakshin10c}.  

Our isolated constant mass planet model is based on the toy model of
\cite{Nayakshin11b} who assumed that the young planet follows a Hayashi-like
track during its early evolution \citep{GraboskeEtal75}. The model gives the
radius of the planet, $r_p(t)$, at time $t$, as
\begin{equation}
r_p^3(t) = {r_{2}^3 \over 1 + A r_{2}^3 t}\;,
\label{re_t}
\end{equation}
where $r_2$ is the planet's radius at the time of the second collapse, and $A
= 24 \pi \sigma_B T_{\rm eff}^4/(GM_p^2)$, where the effective temperature of
the planet, $T_{\rm eff}$ is a free parameter that is fixed for a given
simulation. To accommodate a range of possible cooling rates,
\cite{Nayakshin11b} explored models of contracting planets with three values
of the effective temperature, 500, 1000 and 2000 K, but we test only the mid
range model here.

The initial radius of the planet, $r_2$, right after molecular hydrogen
dissociation is also uncertain, with results depending on dust opacity,
internal energy liberation by a possibly present solid core, planet rotation,
and the detailed equation of state at high densities
\citep[cf.][]{MarleyEtal07}. Following \cite{Nayakshin11b}, we estimate the
initial (post-collapse) radius of the planet by the energy conservation
argument. We assume that the change in the gravitational potential energy of
the planet, $\approx -GM_p^2/2r_p$, as the planet radius collapses, is mainly
used to dissociate H$_2$ molecules and ionise hydrogen atoms. Molecular
hydrogen is completely dissociated after the collapse, whereas the fraction of
fully ionised hydrogen, $0 \leq X_i \leq 1$, is a free parameter of the model
that encapsulates the uncertainties pointed out above. Going through this
simple logic \citep[cf.][]{Nayakshin11b}, we obtain for the initial radius of
the planet,
\begin{equation}
r_2 \approx {G M_p m_H \over D + 2 X_i {\cal E}} \approx 4.5 R_J {M_p
  \over M_J} {1 \over 1 + 2 X_i{\cal E}/D}\;,
\label{re0_est}
\end{equation}
where $D = 4.5$ eV and ${\cal E}=13.6$ are the dissociation energy of hydrogen
molecules and the ionisation energy of hydrogen atom, respectively.  The
lowest starting density models correspond to $X_i=0$, whereas $X_1=1$ give the
densest possible models. 

In equation \ref{re_t} the time $t$ is counted from the time of the second
collapse, which we take to be the start time of the simulations presented
below. This simple model does not take into account the electron degeneracy
pressure. For low density start models, electrons becomes partially degenerate
only after $t\simgt 10^6$ yrs \citep[e.g.,][]{GraboskeEtal75}. We do not run
our models for that long in the present paper. For the high density start
models, we over-estimate the density of the planet (since the electron
degeneracy pressure is neglected), but this becomes important only in the
innermost regions of the disc, $R\simlt 0.02$ AU, where the planets are likely
to be swallowed by the star anyway.  We also neglect the importance of the
possible solid core inside the planet. We plan to address these issues in
future publications.

In practice, we start our calculations with the planet of a given mass $M_p$
right after the second hydrodynamical collapse. The initial planet's radius,
$r_p(0)=r_2$, is given by equation \ref{re0_est}. When the mass loss from the
planet commences (cf. \S \ref{sec:loss}), the planet's radius is advanced
according to equation \ref{drp_dt}. The right hand side of that equation
is the sum of the term due to the planet's mass loss and the radiative cooling
term calculated at a constant planet mass. For our toy cooling model, we have,
\begin{equation}
{1\over \tau_c} = \left({d \ln r_p \over dt}\right)_{\dot M_p=0} = - {1\over
  3}\; A r_p^3 \;,
\label{drp_dtrad}
\end{equation}
where $A= 24 \pi \sigma_B T_{\rm eff}^4/(GM_p(t)^2)$, i.e., calculated with
the evolved $M_p(t)$. For reference, 
\begin{equation}
|\tau_c| = 1.8\times 10^5\mbox{ yr}\; \left({M_p \over 10
  M_J}\right)^2 \left({0.01\mbox{ AU}\over r_p}\right)^3
\left({1000\mbox{ K}\over T_{\rm eff}}\right)^4\;.
\label{tc_num}
\end{equation}

\subsection{Parameter $f_a$ and the gap in the disc}\label{sec:fa_gap}

Our analytical study showed that the outcome of the tidal disruption of a
planet strongly depends on the parameter $f_a$ that describes the fraction of
the specific angular momentum lost through L1 but then regained by the planet
due to gravitational torques from the inner disc (cf. equation \ref{dadt0}).
In numerical experiments below, these torques are calculated
self-consistently, so $f_a$ is neither needed nor introduced. But it may be
quite beneficial to know what an effective value of $f_a$ is for a given
numerical situation to be able to appeal to the analytical insights.

The simplest situation here is when the planet carves out a deep gap in the
disc that stops any significant gas flows across the gap. The outer disc can then
be viewed as a source of an external torque onto the planet, and the situation
is thus quite analogous to that in a semi-detached stellar binary system. In a
binary system, in the conservative mass and angular momentum exchange
scenario, the disc transfers mass in and angular momentum out, giving the
excess specific angular momentum to the secondary. In our problem, if the star
accretes gas with the specific angular momentum of the last circular orbit
around the star (which we set for simplicity to $R=R_*$), $\sqrt{GM_* R_*}$,
and the planet recovers the rest of the specific angular momentum lost through
the L1 point, then
\begin{equation}
f_a = 1 - (R_*/a)^{1/2}\;.
\label{fa_num}
\end{equation}
If $a \gg R_*$, then $f_a \approx 1$.

Now, when the gap is partially or completely closed, gas is able to flow past
the planet, taking with it the excess angular momentum as well. While details
clearly depend on how deep the gap is, we expect that $f_a$ becomes smaller as
the gap closes, and in fact goes to zero in the limit of no gap.

According to \cite{TakeuchiEtal96}, the planet opens an annual gap in the disc
if $M_p$ exceeds
\begin{equation}
{M_p \over M_*} > 2 \alpha^{1/2} \left({H \over R}\right)^2\;.
\label{gap}
\end{equation}
As $H/R \sim 0.1-0.2$, the estimated gap opening mass varies from a
sub-Jupiter to a few Jupiter masses depending on the $\alpha$-parameter of the
disc. Obviously, various degrees of stellar irradiation and different opacity
laws can affect the gap opening through increasing or decreasing the
geometrical aspect ratio $H/R$.

\section{Tests of planetary destruction}\label{sec:CD}

We now present a series of controlled numerical experiments. To focus our
attention on the planet only, we do not yet introduce an outer disc.  Instead,
we express the effects of the outer disc by a fixed torque,
\begin{equation}
\left({d \ln a\over dt}\right)_{\rm ext} =  {1 \over \tau_{\rm e}}\;,
\label{dadt_ext}
\end{equation}
where $\tau_{\rm e} = - 250$ yrs for the tests below. Except for the
  extreme mass ratio, $M_p/M_*$, the setting of the problem is exactly
  analogous to that of mass transfer in stellar binaries, where the external
  torque might be due to the gravitational radiation or magnetic breaking. 
Such a setup facilitates a straight forward comparison with the analytical
theory for the planet migration and Roche lobe overflow developed earlier.
 However, we shall find ``new'' behaviour compared with stellar binaries
  motivated by the disc gap closing (something that never happens for stellar
  binaries). More self-consistent simulations, those without an external
torque and where the outer disc torques the planet in, are to follow in \S
\ref{sec:full}.

In this section, the planet initial mass is $M_p = 10 M_J$, the initial
planet-star separation is $a_0 = 0.25$ AU, the initial planet's radius, $R_p =
0.021 \mbox{AU} \approx 44 R_{\rm J}$, obtained from equation \ref{re0_est}
with $X_i=0$. The effective temperature of the planet is set to $T_{\rm eff} =
1000$~K. We do not allow the planet's mass to fall below $M_{\rm core} =
5\mearth$, assuming that there is a solid core of that mass inside the
planet. This particular assumption only affects the results at the very end of
the tidal disruption of the planet. We use Thompson scattering opacity for the
disc for simplicity (we shall use a more realistic one for the full disc
simulations later on).  The inner radius of the computational domain, assumed
to equal the stellar radius, is $R_{\rm in} = 0.01$ AU, and the outer radius
is $R_{\rm out} = 4.2$ AU (which is sufficient for these tests as we are only
interested in the inner regions). The number of grid zones is $N_r = 500$,
unless otherwise stated.

Table 1 shows the list of the simulations presented in this paper, most
important parameter values, respective figure numbers and also the most
notable behaviour patterns of the disc or the planet.

\subsection{Simulation Ext1}\label{sec:ext1}

The first run presented, labelled Ext1, is for a disc viscosity parameter
$\alpha = 0.02$ and the planet radius-mass exponent of $\zeta_p = -1/3$.

\subsubsection{Evolution of the planet}\label{sec:eplanet}

\begin{figure}
\centerline{\psfig{file=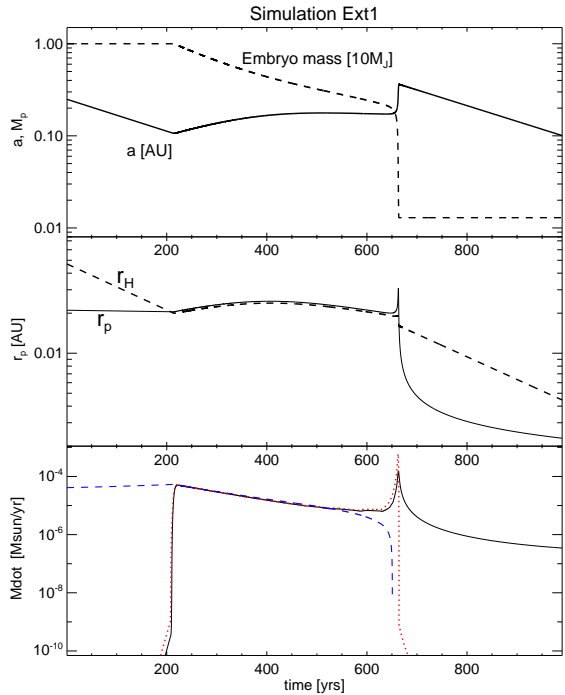,width=0.5\textwidth,angle=0}}
\caption{Evolution of a massive gas planet tidally disrupted near its parent
  star. The top panel shows the radial location and planet's mass versus
  time, as labelled. The middle panel shows the planet's radius, $r_p$, and
  its Hills radius, $r_H$. The bottom panel shows the accretion rate onto the
  star, the planet mass loss rate and the theoretical prediction for that rate
  (equation \ref{dotm_eq}) with the solid black, red dotted and blue dashed
  curves, respectively. The theoretical prediction works exceedingly well in
  the region where the planet fills its Roche lobe and the gap in the disc is
  opened (see \S \ref{sec:eplanet}).}
\label{fig:CD_hist}
\end{figure}

Figure \ref{fig:CD_hist} shows the most interesting characteristics of this
simulation. The solid line in the top panel shows the planet's location, $a$,
as a function of time, in units of AU. The dashed curve shows the planet mass
in units of $10 M_J$. The solid curve in the middle panel shows the radius of
the planet, $R_p$, whereas the dashed curve shows the Hill's radius. The
bottom panel of Fig. \ref{fig:CD_hist} presents three mass loss/gain rates:
the accretion rate onto the star, the absolute value of the planet mass loss
rate, $|\dot M_p|$, and the analytical prediction for that mass loss rate given
by equation \ref{dotm_eq}, shown with the solid, the red dotted and the blue
dashed curves, respectively.

The general outcome of the simulation is that the planet spirals in for about
200 years with zero mass loss, until it fills its Roche radius and starts
losing mass. The mass lost by the planet is transferred into the disc
interior to it. A quasi-steady state is established in which the mass loss
by the planet is exactly matched by the mass accreted onto the surface of the
star. The inner disc transfers its excess angular momentum back to the planet,
producing an outward directed torque on the latter. This reverses the
direction of planet's migration. Losing mass, it is migrating outward, as
predicted. The quasi-steady state mass loss cannot last forever as the
planet's mass is finite, and eventually the mass loss runs away, destroying
most of the gas component of the planet rapidly (see the spike at around 660
yrs). Only $\sim 0.12 M_J$ of the gaseous planet survive the disruption in
this simple model (see \S \ref{sec:spike} below).

The blue dashed curve in the lower panel of Fig. \ref{fig:CD_hist} shows the
predicted mass loss rate given by \ref{dotm_eq} with $f_a$ given by equation
\ref{fa_num}. We observe that the analytical prediction provides an excellent
description of the numerical results starting from the time when the planet
fills its Roche lobe, e.g., from time $t\approx 200$ yrs to $t\approx 550$
yrs. During this time the mass loss rate from the planet is matched by the
accretion rate of the star. The second marker of this quasi-equilibrium is the
fact that the rate of planet's expansion/contraction due to mass loss and
radiative cooling is exactly matched by the corresponding change in the Hill's
radius, so that the two are approximately equal (cf. the middle panel of
Figure \ref{fig:CD_hist}). Note that the radius of the planet increases from
$t \approx 200$ to $\approx 400$ yrs, and then decreases until the sharp
spike. This change in behaviour is explained by the fact that initially the
planet's radius evolution is controlled by the mass loss, with cooling being
negligible. Thus the planet initially expands. After $\approx 400$ yrs, when
the planet loses about half of its original mass, the radiative cooling
becomes dominant in controlling the planet's radius evolution. The planet then
contracts until the accretion rate spike.

The quasi-equilibrium state of the planet-star system is analogous to that of
a semi-detached stellar binary system where the lower mass secondary transfers
its mass to the primary. However, after $t\approx 550$ yrs the planet's mass
loss rate stops decreasing with time, flattens off and then increases. The
accretion rate onto the star follows the trend, but lagging behind the
planet's mass loss. As we shall see below, this is not a time-delay effect but
rather is a signature of a very important effect that does not occur in
stellar binaries: the closing of the gap in the disc, due to which the inner disc material may
flow past the planet. As the result of this, the planet suffers a nearly
catastrophic disruption at $t\approx 660$ yrs.

\subsubsection{Evolution of the disc}\label{sec:edisc}

\begin{figure}
\centerline{\psfig{file=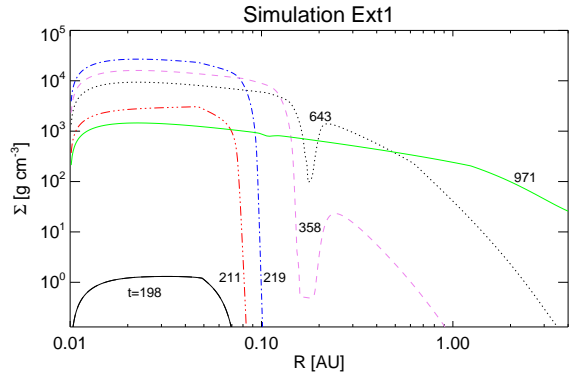,width=0.5\textwidth,angle=0}}
\caption{The disc surface density for several different times, as marked on
  the Figure. The evolution of the disc can be summarised as following: (a)
  the inner disc is inflated with material from the disrupted planet; (b) the
  disc pushes the planet outward; (c) the gap is partially closed,
leading to a runaway in planet's mass loss rate; (d) the disc overflows the
remaining low mass planet. The evolution after that is that of a ``standard''
accretion flow unperturbed by the planet.}
\label{fig:CD_disc}
\end{figure}

Figure \ref{fig:CD_disc} shows several snapshots of the disc surface density
versus radius covering the most interesting moments of this numerical
experiment. The curves are marked by the respective time $t$ shown right next
to them. As stated earlier, initially the surface density profile is
everywhere zero. At $t\approx 200$ yrs (see the middle panel of Figure
\ref{fig:CD_hist}), the planet eventually fills its Roche lobe and starts
transferring mass into the inner disc, filling it out. The first curve shown
in Figure \ref{fig:CD_disc} corresponds to time $t=198$ yrs (black solid
curve), when the planet disruption just begins. The planet at that moment is
at $a = 0.12$ AU. The material lost through the L1 point is circularised and
is deposited at $R_{\rm dep}\approx 0.06$ AU (note the break in the black
curve in the Figure), and then spreads viscously in both directions.  At small
radii the material eventually starts accreting onto the star, whereas at
larger radii the torques from the planet prevent it from spreading outwards
further. During this time the mass loss rate from the planet exceeds the
accretion rate onto the star (the rising part of the red dotted curve at
 $t\approx 200$ yrs in the bottom panel of Figure \ref{fig:CD_hist}). The
inner disc is thus being filled in by the material lost from the planet.

The swelling of the inner disc cannot go on forever, however. When the mass of
the disc becomes significant enough ($t=211$ and $t=219$ yrs curves in Figure
\ref{fig:CD_disc}), the gravitational torque of the inner disc starts to
affect the radial migration of the planet. The inner disc torque in fact
exceeds that of the externally applied torque and the planet reverses its
direction of migration. Migrating outward, it is now able to settle into 
the self-regulated quasi-steady regime discussed earlier.

This amiable quasi-steady state evolution of the planet and the inner disc
continues until time $t\approx 500-600$ yrs or so. By that time, the mass of
the planet drops to only $\sim 3 M_J$. This severely undercuts the ability of
the planet to keep the gap open (cf. equation \ref{gap}). This is why the gap
starts to be partially filled by gas, and the material from the inner disc
starts to leak into the outer one (cf. the $t=358$ yrs curve in Figure
\ref{fig:CD_disc}). By $t=643$ yrs curve, the gap is seriously
compromised. The last of the curves, $t=971$ yrs, shows that in the end the
gap is completely erased and the disc spreads viscously outward even more. By
that time the planet has lost most of its mass and the disc is hardly aware of
its presence (even though it consists entirely from the material formerly
belonging to the planet in this simulation).

\subsubsection{The late spike/final destruction of the
  planet}\label{sec:spike}

The partial gap closing at $t\sim 600$ yrs has catastrophic consequences for
the planet. As the inner disc material diffuses outward through the gap, past
the planet's orbit, less of the inner disc torque is passed on to the planet. In
terms of the analytical parameterisation of \S \ref{sec:analytical}, the
fraction of the angular momentum passed back to the planet, $f_a$, drops.
Since $\zeta_p - \zeta_H = 2(f_a-1/3)$ for $\gamma=5/3$ planet, the mass
transfer rate from the planet becomes unstable when $f_a$ drops below
$1/3$. The mass loss rate runs away exponentially at time $t\approx 660$
yrs. Physically, the planet's radius increases more rapidly due to the mass
loss than the Hill's radius could increase since the outward radial migration
is compromised by the leaking gap. As can be seen from the bottom panel of
figure \ref{fig:CD_hist}, during this exponential runaway, the protostellar
accretion rate does not manage to keep up with the rate at which the matter is
deposited in it. This is not surprising as some of the matter flows outward
through the gap.

The runaway planet destruction via Roche lobe overflow is bound to end in one
of the two ways. One is a complete unbinding of the gaseous envelope that
would leave only a rocky core (which is present for this simulation by
assumption). On the other hand, if the cooling time of the planet becomes very
short at low masses, the planet may go into a ``runaway contraction'' phase
instead, where it contracts faster than the Hill's radius expands. This second
outcome is indeed what happens in simulation Ext1. The end mass of the planet
is $0.12 M_J$, including the assumed $5\mearth$ of the rocky core. 

 We emphasise that the final mass of the planet strongly depends on its
  internal structure, and so is very model dependent. In our simple constant
  $T_{\rm eff} =1000$ K cooling prescription, contraction time scale becomes
  very short at low planet's masses. This shuts off the mass loss at the end
  of simulation Ext1. A more realistic cooling model would probably lead to
  longer cooling times. On the other hand, the solid core may have created its
  own dense gas ``atmosphere'' strongly bound to the core, in analogy to the
  nucleated core instability model for giant planet formation
  \citep[e.g.,][]{Mizuno80,PollackEtal96}, as argued by
  \cite{Nayakshin10b}. This atmosphere is not modelled here, and may remain
  bound to the core. Clearly, better models of the internal structure of the
  planets must be used to address the issue of the post-disruption mass.

\subsection{Importance of gap opening: simulation Ext2}\label{sec:gap}

As we have seen in simulation Ext1, when the gap closes, $f_a$ drops since the
matter and angular momentum flow from the inner disc into the outer disc, past
the planet. This reduces the rate of the outward migration of the planet and
may destabilise the mass loss by the planet, leading to its runaway
disruption.  In view of the gap opening criterion (equation \ref{gap}), then,
the outcome of the planetary tidal disruption depends on the disc viscosity,
the planet mass, and the outer disc torque. The latter controls the accretion
rate in the inner disc (through the equilibrium mass loss condition in
equation \ref{dotm_ext}) which in its turn regulates the geometrical disc
aspect ratio, $H/R$.

\begin{figure}
\centerline{\psfig{file=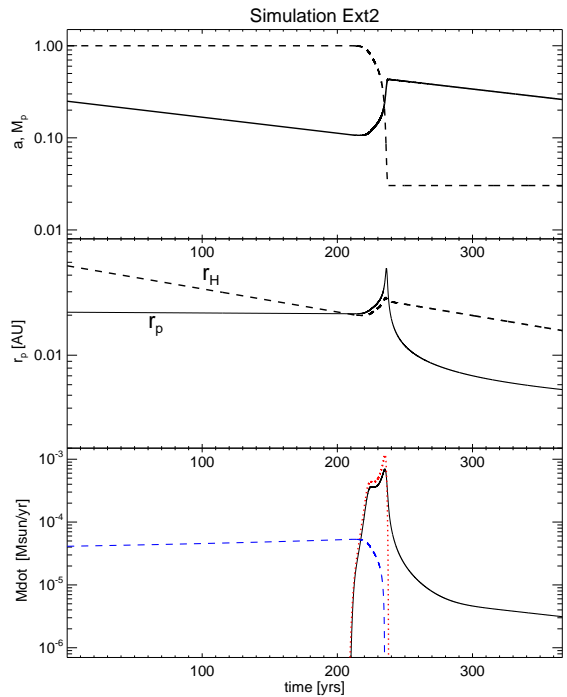,width=0.5\textwidth,angle=0}}
\caption{Same as Fig. \ref{fig:CD_hist} but for simulation Ext2 with a higher
  disc viscosity parameter, $\alpha=0.1$. Due to this, the disc is hotter, and
  the planet is unable to keep the gap opened. As the result, the material
  from the inner disc is able to flow past the planet. The reduced angular
  momentum feedback on the planet destabilises the mass transfer process. Most
  of the planet's gaseous inventory is completely disrupted on a very short
  time scale.}
\label{fig:CD_hist2}
\end{figure}

To explore this issue, we run one further simulation, labelled Ext2, entirely
identical to Ext1 except that the viscosity parameter $\alpha$ is increased to
$\alpha=0.1$. Figure \ref{fig:CD_hist2} shows the resulting time evolution of
the planet in the same format as Figure \ref{fig:CD_hist}. The outcome of the
higher $\alpha$ simulation is markedly different from that of its lower
$\alpha$ counterpart. There is no quasi-steady state plateau in the accretion
rate corresponding to an equilibrium tidal evaporation of the planet. Analysis
of the disc surface density evolution (not shown here for brevity) for
simulation Ext2 shows that there is only a depression in the surface density
profile rather than a deep gap. As explained earlier, a partially opened gap
lowers the effective $f_a$. The planet does not move outwards quickly enough,
and the expansion of the planet (recall that $R_p\propto M_p^{-1/3}$)
increases the mass loss rate further, until the planet is destroyed in a
runaway fashion.


\subsection{Importance of the mass--radius relation: simulation
  Ext3}\label{sec:ext3} 

\begin{figure}
\centerline{\psfig{file=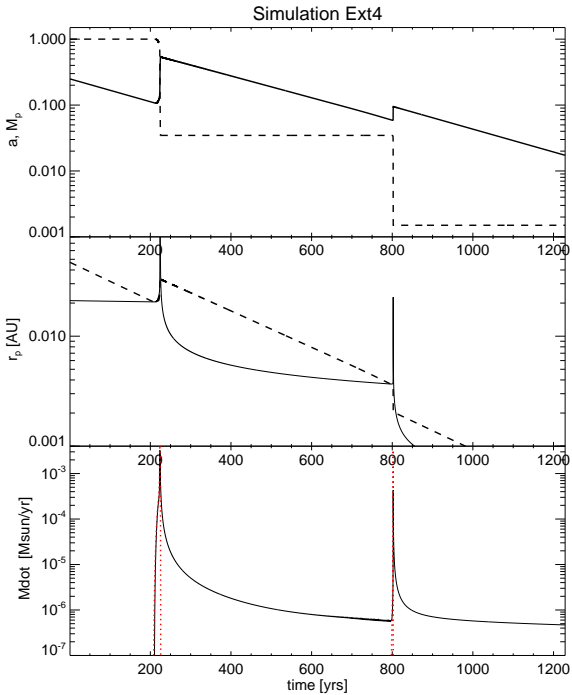,width=0.5\textwidth,angle=0}}
\caption{Same as Fig. \ref{fig:CD_hist} but for simulation Ext3. Here the
  mass--radius relation of the planet is steeper, $r_p\propto M_p^{-1}$.  The
  mass-loss from the planet fuels too fast an expansion of $r_p$, so that the
  difference $r_p -r_H$ ``runs away'', and so does the planet mass loss
  rate. The planet loses most of its mass on a very short time scale in the
  first very luminous burst. The remaining gas envelope is unbound in the
  second disruption when the planet migrates inward even closer.}
\label{fig:Ext3}
\end{figure}

As a final numerical experiment in a series with the fixed external torque
timescale, we present simulation Ext3, in which the mass--radius relation for
the planet is steeper, $r_p\propto M_p^{-1}$. That is, $\zeta_p=-1$ rather
than $-1/3$ for simulations Ext1 and Ext2. Figure \ref{fig:Ext3} shows the
outcome of this simulation, demonstrating that there are actually two mass
loss episodes and both are unstable.

The stability of the mass transfer depends (see \S \ref{sec:quasi_steady}) on
the difference $\zeta_p-\zeta_H = 2f_a -4/3$ in this case. If the gap is
impenetrable to the gas from the inner disc, and the material lost by the
planet is {\em instantaneously} gained by the star, then $f_a = 1 -
\sqrt{R_{\star}/a}$, as explained in \S \ref{sec:fa_gap}.  The planet is first
disrupted at $R\approx 0.11$ AU in this experiment, so $f_a\approx 0.7$, and
$\zeta_p-\zeta_H = 2f_a -4/3 \approx 0.1$, i.e., small but marginally
positive. However, there is a small lag between the mass loss rate of the
planet and the accretion rate, which reduces the effective value of $f_a$
below the above (theoretically maximum) estimate. This is probably the reason
for the mass transfer being unstable in this simulation.

The first disruption episode is not final in this simulation. As the planet
loses all but $\sim 0.3 M_J$ of its gaseous mass (which corresponds to only
3\% of its initial mass), the radiative cooling accelerates, and $r_p$ shrinks
rapidly. At the same time, the planet migrates outward, so the Hill's radius
increases. This allows the remaining Saturn-mass planet to shut down the mass
loss for a while. However, the inner disc drains onto the star, while the
``external torque'' in our fixed $\tau_e$ model continues to push the planet
inward. So the planet resumes its inward migration. By the time $t= 800$ yrs
it finds itself at $R = 0.06$ AU, where it fills its Roche lobe for the second
time. A second outburst, physically but not quantitatively similar to the
first one, results. The remaining reservoir of the gas is lost. The ``naked''
solid core then continues to migrate inward as prescribed by the fixed
$\tau_e$ model.

\section{Disruption of planets in realistic discs}\label{sec:full}

\subsection{Parameters and initial conditions}

We now consider a more realistic set of calculations in which we do not
introduce an artificial external torque. Instead, we start with an accretion
disc filling in the computational domain. The torque
parametrized in \S \ref{sec:CD} by the timescale $\tau_{\rm e}$ is now
self-consistently computed from the time-dependent disc structure. The disc is
initialised with a surface density profile commonly used in the literature
\citep[e.g.,][]{MatuyamaEtal03,AlexanderEtal06} in studies of protostellar
disc evolution:
\begin{equation}
\Sigma_0(R) = {A_m \over R} \left(1 - \sqrt{R_{\rm in}\over R}\;\right)\;\exp\left[-{R\over R_0}\right]\;,
\label{sigma0}
\end{equation}
where $R_0$ is the disc length-scale, set at $R_0 = 5$ AU unless stated
otherwise, and $A_m$ is a normalisation constant chosen so that the disc
contains a given initial mass $M_{d} = 2\pi \int_{R_{\rm in}}^{R_{\rm out}} R
dR \Sigma_0(R)$. In hindsight, the exact shape of the initial surface density
profile does not appear very important, as long as one samples all the
interesting parameter space by varying other parameters of the problem, such
as the disc mass $M_{d}$ and the viscosity parameter $\alpha$. We use the
opacity of \cite{ZhuEtal09} instead of Thompson electron scattering opacity,
used in tests Ext1--Ext3. The computational domain is modelled by $N_r=500$ to
700 radial zones. We also use $R_{\rm in} = 0.02$ AU for the inner boundary of
the disc, instead of 0.01 AU as used in tests Ext1--Ext3. The outer boundary
of the computational domain is set to $R_{\rm out} = 25$ AU for the tests
below, or otherwise explicitly stated. We use a reflecting boundary condition
at $R_{\rm out}$. This choice is of no particular importance for the results
as there is little gas mass at large $R$ and the most interesting effects take
place at $R\simlt 1$ AU. In addition, the viscous time at $R_{\rm out}$ is
much longer than the duration of the simulations below. The initial mass of
the planet is set at $M_p=10 M_J$.

\subsection{Simulation Sim1}

Simulation Sim1 is done with the following parameters: viscosity parameter
$\alpha= 0.002$, $M_d= 20 M_J$, $X_i = 0$, $T_{\rm eff} = 1000$ K, $\zeta_p =-
1/3$. The initial location of the planet, $a_0 = 0.13$ AU, is chosen to be
such that the Roche radius $r_H$ is just slightly larger than the planet
radius, $r_p$. This is done for computational expediency. Tests Sim3 and Sim4 below
are started with larger value for $a_0$.

\begin{figure}
\centerline{\psfig{file=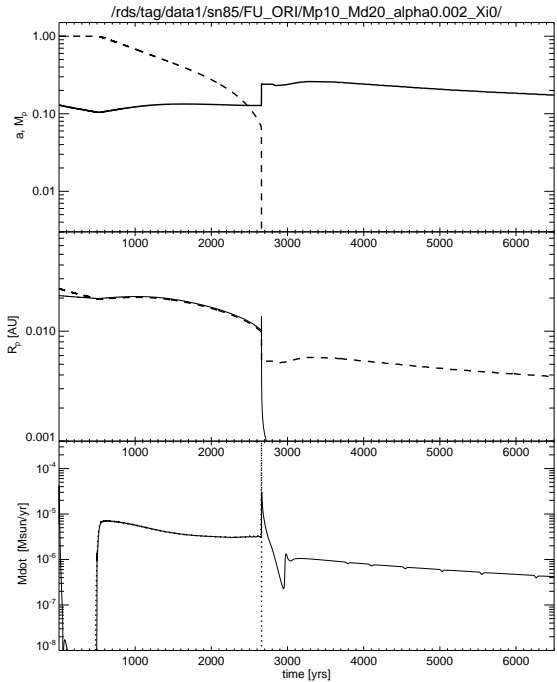,width=0.5\textwidth,angle=0}}
\caption{Same as Fig. \ref{fig:CD_hist} but for simulation Sim1 that models
  the disc torques self-consistently.}
\label{fig:hist_Sim1}
\end{figure}

Figure \ref{fig:hist_Sim1} shows the resulting evolution of the planet and the
accretion rate onto the star. In brief, the time evolution of the system can
be described as the following stages: (i) From $t=0$ to $t\approx 100$
yrs. The evacuation of the disc interior of the planet; (ii) $t\approx 100$ to
400 yrs. An inward migration of the planet and then commencement of the Roche
lobe overflow, which re-fills the inner disc; (iii) $t\approx 400$ to 2600
yrs; the quasi steady state outward migration of the planet as the result of
the inner disc torque exceeding that of the outer disc torque. The planet
loses most of its mass during this time, evolving from $10 M_J$ to only about
$1 M_J$. (iv) $t\approx 2660$ yrs. The gap is partially closed. This reduces
the inner disc torque and destabilises the quasi-equilibrium condition ($r_p
\approx r_H$). The planet mass loss runs away since $r_p$ expands faster than
$r_H$ could increase due to the outward migration. This leads to the ``final
destruction'' spike in the planet mass loss rate (the dotted curve in the
bottom panel of figure \ref{fig:hist_Sim1}). The planet loses almost all of its
gaseous mass in this particular simulation. (v) $t>2660$ yrs, the
``post-planet'' disc evolution. With the planet now all but destroyed (the
remaining solid core is dynamically unimportant for the disc), the disc
evolves independently of the planet.

\begin{figure*}
\centerline{\psfig{file=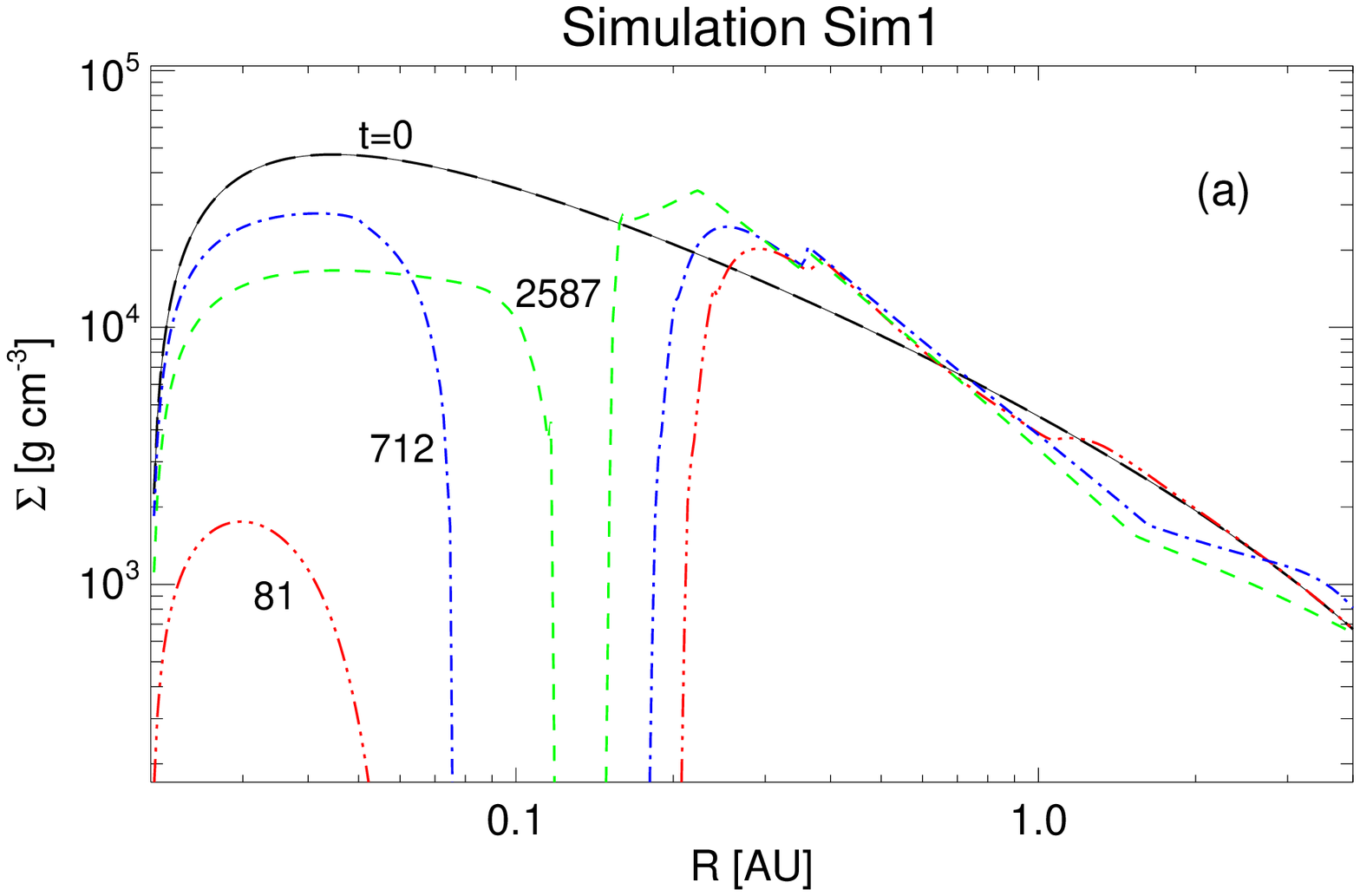,width=0.5\textwidth,angle=0}
\psfig{file=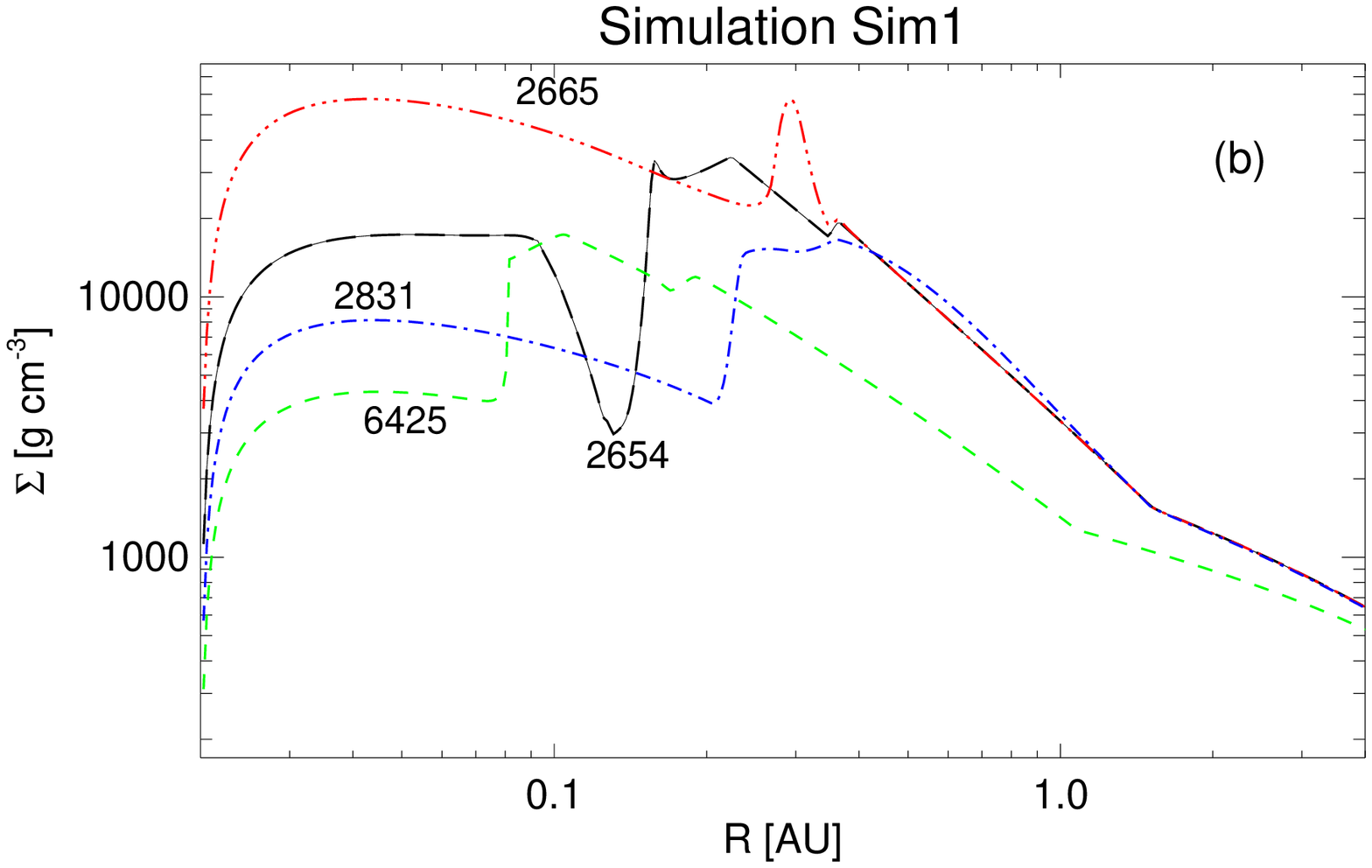,width=0.5\textwidth,angle=0}}
\caption{The disc surface density profile for several snapshots from the
  simulation Sim1. The times of the snapshots are labelled on the figure next
  to the respective curves. (a) The early evolution of the system. The $t=0$
  initial condition neglects the presence of the planet, but the planet opens
  a gap in the disc very quickly. The inner disc then empties out onto the
  star. The outer disc then ``pushes'' the planet closer in where it fills its
  Roche lobe and starts transferring mass to the inner disc. The inner disc then
  refills with the {\em planet} material. Note that the gap becomes narrower
  as the planet loses mass. (b) Later evolution of the disc, centred around
  the spike at $t=2660$ yrs (see Figs. \ref{fig:hist_Sim1} and
  \ref{fig:histZoom_Sim1}). The planet is rapidly dispersed when the mass loss
  rate runs away due to the partially closed gap. An ionisation front
  propagates outward, but stalls quickly. After the giant planet is destroyed,
  the disc accretes onto the star as if there were no planet.}
\label{fig:S1_disc1}
\end{figure*}

We now discuss this evolution in greater depth. Figure \ref{fig:S1_disc1}
shows the disc surface density profile at key epochs. Panel (a) concentrates
on the earlier times, when the planet is still massive enough to keep the gap
opened. The solid curve shows the initial $\Sigma(R)$. The next curve at
$t=81$ yrs shows a greatly reduced surface density in the disc interior to the
location of the planet. It also demonstrates that the planet pushes the outer
disc back, creating a bump in $\Sigma(R)$ just behind the gap. After the inner
disc disappears (accretes onto the star), the outer disc pushes the planet
closer to the star. Roche lobe overflow sets in, and the inner disc is
refilled by the planet's material. The $t=712$ curve in Figure
\ref{fig:S1_disc1}a shows that it is filled almost to the same values of
$\Sigma$ as at $t=0$. The planet now migrates outward, and the gap becomes
less wide since the planet's mass is reduced. This description (and Figure
\ref{fig:S1_disc1}a) covers simulation stages (i) to (iii), as defined above.

Figure \ref{fig:S1_disc1}b shows what happens when the planet's mass is too
small to keep the gap opened. The gap is significantly compromised by the time
of the black solid curve, $t=2654$ yrs. The gap is completely closed, and the
planet's mass is dispersed and assimilated into the inner disc in the next
$\sim 20$ yrs (see the $t=2665$ yrs curve). The mass released by the planet
streams both inwards -- onto the star -- and outwards (note the spike in the
red $t=2665$ yrs curve in the figure).

\begin{figure}
\centerline{\psfig{file=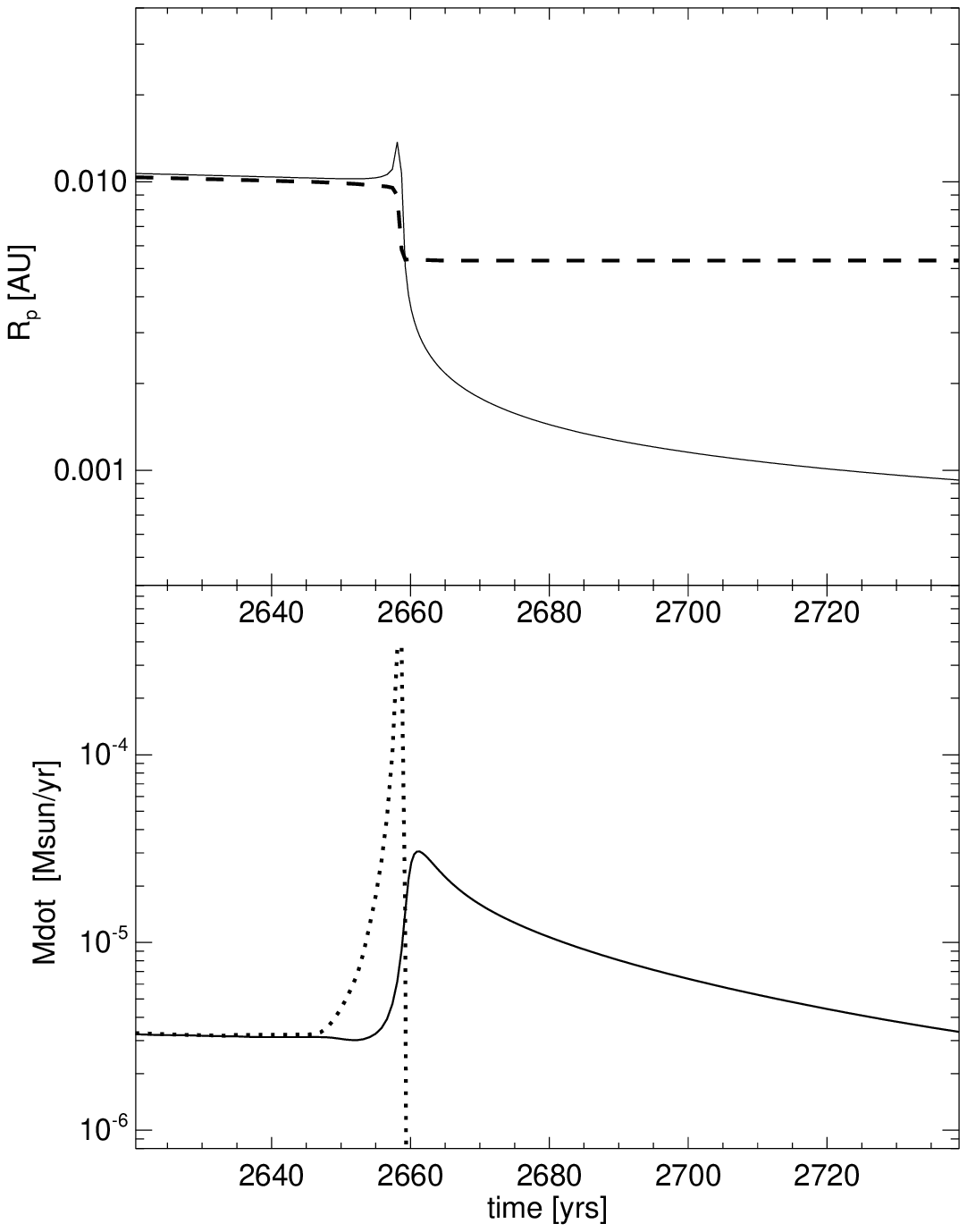,width=0.5\textwidth,angle=0}}
\caption{A zoom in on the evolution of the planet's radius, and the Hills
  radius (top panel), and the planet mass loss rate and the star's accretion
  rate (the bottom panel) for simulation Sim1.}
\label{fig:histZoom_Sim1}
\end{figure}

Figure \ref{fig:histZoom_Sim1} zooms in onto the time interval around the
``final destruction'' of the planet, e.g., around the gap closing, showing the
planet's radius and Hills radius in the top panel, and the accretion rate
onto the star (solid) and the planet's mass loss rate (dotted) in the bottom
panel. The planet's destruction in this simulation ends when only $\approx
0.1\mearth$ of gas remains bound to the $5\mearth$ solid core we assumed in
this simulation. Given our toy cooling model (fixed $T_{\rm eff}$), this
amount of mass cools very quickly, allowing the radius to contract very fast
and avoid a complete unbinding of all of the gaseous mass. It is clear that
this outcome is completely dependent on the cooling model, which becomes
unrealistic at this point. Nevertheless, this demonstrates an important
point. If radiative cooling becomes sufficiently rapid as the planet loses
mass, the planet (the gas atmosphere of the solid core, actually) may contract
below the Hill's radius before the evaporation of all of the gas. The end
result of this is then a solid core plus a gas envelope planet.

We see from Figure \ref{fig:histZoom_Sim1} that most of the material lost by
the planet in the ``last disruption'' outburst is accreted by the star within
half a century or so (see also the curve $t=2831$ yrs in
fig. \ref{fig:S1_disc1}b). The disc then evolves independently of the planet,
accreting onto the star. The sharp transitions in $\Sigma(R)$ in Figure
\ref{fig:S1_disc1}b at late times ($t=6425$ yrs, for example) are caused by
the strong opacity jumps where Hydrogen becomes partially ionised.

\subsection{Sim1.X: More compact start models}\label{sec:Sim3}

So far we explored only the low density start models, introduced by setting
the parameter $X_i = 0$ (cf. equation \ref{re0_est}). This parameter stands
for the fraction of the ionised hydrogen atoms in the planet right after the
second hydrodynamical collapse. In our toy model for the planet this parameter
controls the radius of the planet right after H$_2$ dissociation. The larger
the initial radius of the planet, the farther away from the star can the
planet be disrupted. We now explore how the results depend on the value of
$X_i$, keeping the other parameters of the simulation fixed at same values as
in Sim1 (cf. Table 1). We label the simulations Sim1.X, where $X = 10 \times
X_1$.

\begin{figure}
\centerline{\psfig{file=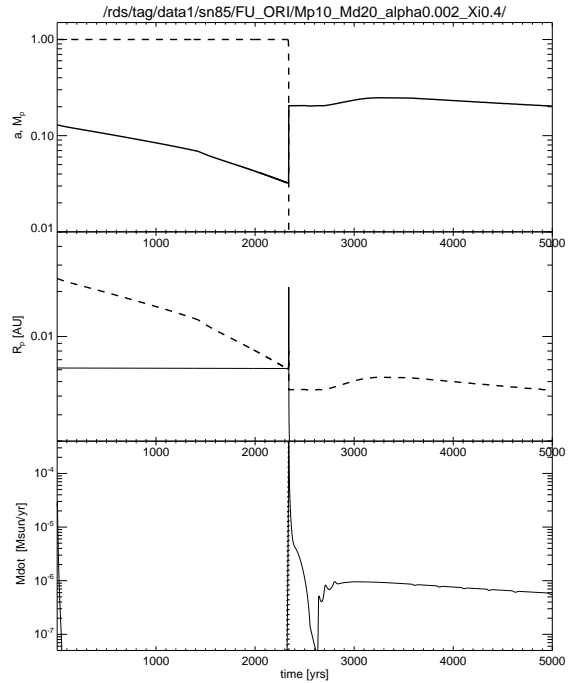,width=0.5\textwidth,angle=0}}
\caption{Same as Fig. \ref{fig:hist_Sim1} but for simulation Sim1.4 which
  differs from Sim1 by a higher hydrogen ionisation parameter, $X_i=0.4$. The
  planet is thus about 3 times more compact, thus being disrupted closer in to
  the protostar. The disruption is abrupt and non steady because
  $\zeta_p-\zeta_H <0$ for this simulation (see \S \ref{sec:Sim3} for
  detail).}
\label{fig:hist_Xi0.4}
\end{figure}

Figure \ref{fig:hist_Xi0.4} shows the outcome of simulation Sim1.4, with $X_i
= 0.4$. Since the initial radius of the planet is less than one third that of
simulation Sim1 (compare the solid curves in the middle panels of
Figs. \ref{fig:hist_Xi0.4} and \ref{fig:hist_Sim1}), the planet is able to
migrate farther in before it fills its Roche lobe. The planet is located at
$R\sim 0.035$ AU at time $t\approx 2300$ yrs when it fills its Roche lobe. In
contrast to simulation Sim1, the mass transfer rate is not quasi-steady in
this case. The whole of the planet's envelope is tidally disrupted in a short
Roche lobe overflow event. The main destabilising factor is the decreased
fractional amount of the angular momentum that the planet receives back when
its mass is accreted by the protostar. In particular, $f_a = 1 - \sqrt{R_{\rm
    in}/a} \approx 0.24$ for the present simulation, whereas for Sim1
$f_a\approx 0.55$ due to the more distant disruption location.  For
$\zeta_p=-1/3$, we have $\zeta_p - \zeta_H = 2(f_a - 1/3)$. Therefore, for
simulation Sim1, $\zeta_p - \zeta_H > 0$, whereas for Sim1.4 $\zeta_p -
\zeta_H < 0$, explaining the vastly different outcomes of these simulations.

The mass deposited in the inner disc drives the planet outwards to about 0.2
AU. During this short planet Roche lobe overflow spike our model for the mass
deposition is probably somewhat unrealistic, as the mass may also overflow
through the L2 point, in the outer disc, at these very high outflow rates.

\begin{figure}
\centerline{\psfig{file=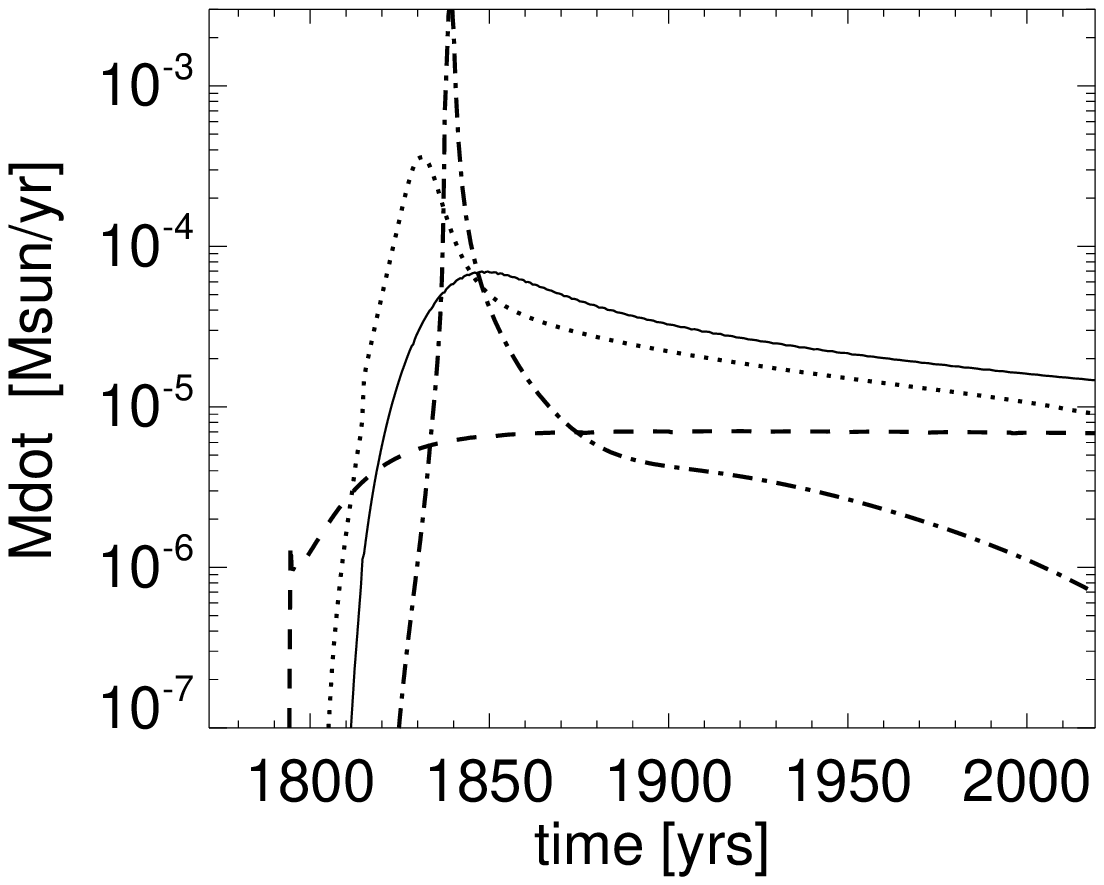,width=0.5\textwidth,angle=0}}
\caption{Comparison of the protostellar accretion rates, $\dot M_*$, around
  the first Roche lobe overflow event for simulations Sim1 and Sim1.X. The
  simulations differ from one another only by the initial ionised hydrogen
  fraction, $X_i=0, 0.2, 0.3$ and 0.4, for the dashed, solid, dotted and
  dash-dot curves, respectively. Note that the more compact the planet is (the
  larger $X_i$ is), the shorter the rise time of the outburst and the larger
  the peak value of $\dot M_*$. The curves were arbitrarily translated along
  the time coordinate to appear on the same figure.}
\label{fig:hist_compXi}
\end{figure}

Figure \ref{fig:hist_compXi} shows the stellar accretion rate evolution near
the tidal disruption point for simulations Sim1, Sim1.2, Sim1.3 and
Sim1.4. The rise times of the light-curves are between a year to ten, but the
more compact the planet is (the larger $X_i$), the steeper the rising and the
falling parts of the curves. This is natural, as the more compact models are
disrupted closer in. The rise time of the light-curves is about the inner disc
viscous time, which becomes shorter at smaller values for deposition radius,
$R$, scaling as $\propto R^{3/2}$ (at a fixed $\alpha$ parameter and $H/R$).

\subsection{Simulation Sim2: higher disc mass}\label{sec:Sim2}

\begin{figure}
\centerline{\psfig{file=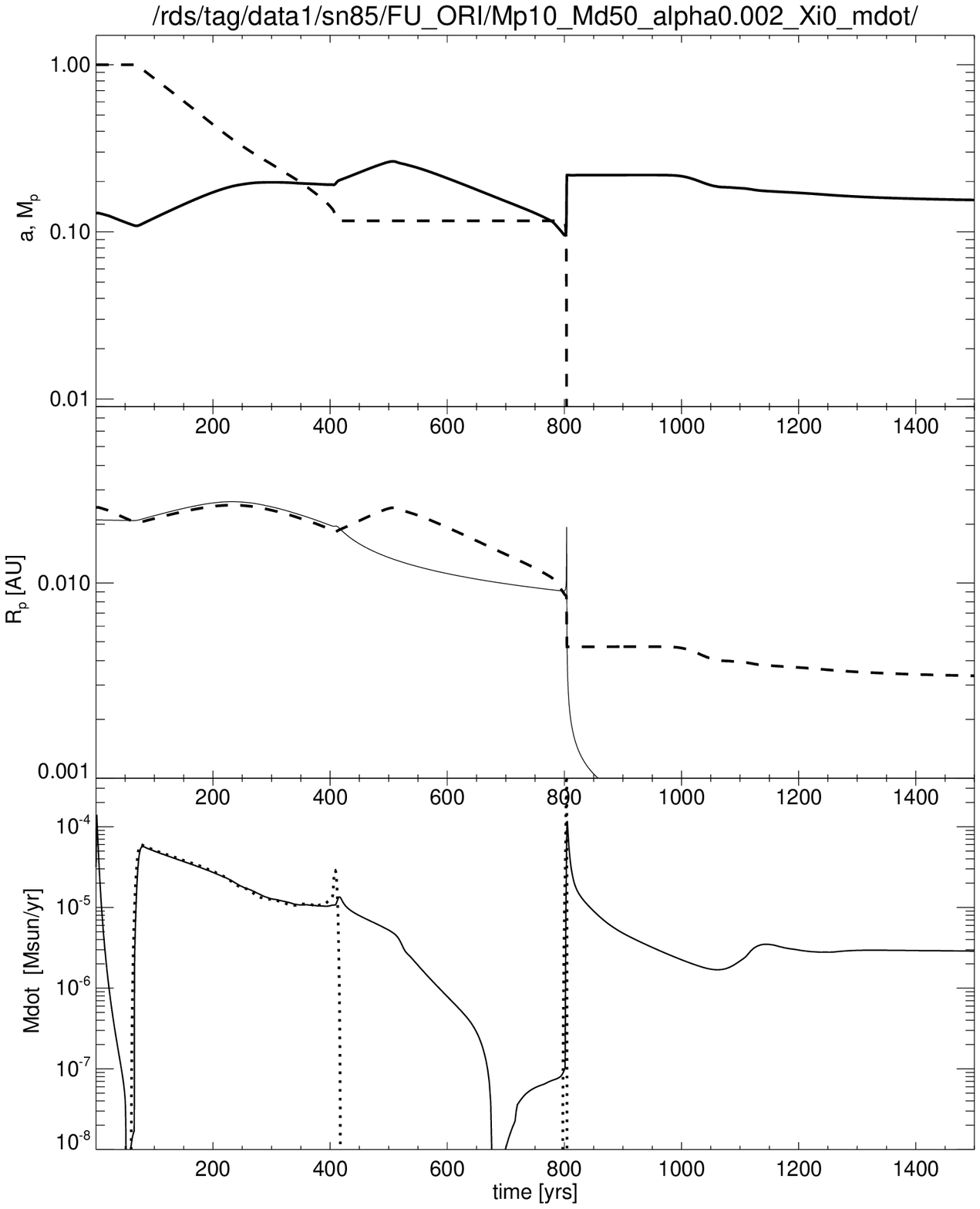,width=0.5\textwidth,angle=0}}
\caption{Same as Fig. \ref{fig:hist_Sim1} but for simulation Sim2. Note that
  the timescales are shorter and the mass loss rates are larger in Sim2 than
  they are in Sim1.}
\label{fig:hist_Sim2}
\end{figure}

We now present simulation Sim2, which is identical to Sim1 except for a higher
initial accretion disc mass, $M_d=50 M_J$. Figure \ref{fig:hist_Sim2} shows
the main results of this numerical experiment. Since the disc accretion rate
is higher, the torques acting on the planet are higher too. Therefore the time
scales for planet migration in this simulation are shorter than in Sim1. In
terms of our analytical approach to the problem, the external disc torque
timescale, $|\tau_e|$, is shorter in Sim2 than in Sim1.

As before, the inner disc is first emptied onto the star. The planet fills its
Roche lobe at around $t=100$ yrs, starting a very large accretion
event. Note that the maximum accretion rate reached in this simulation at the
first Roche lobe overflow event is nearly $10^{-4} \msun$~yr$^{-1}$, almost an
order of magnitude higher than in simulation Sim1, which is explainable
by the larger disc torques onto the planet. 

During the quasi-steady Roche lobe overflow stage, the planet migrates
outward, as before. Interestingly, by the time it reaches $1 M_J$ mass, the
planet has migrated outward to $R\approx 0.2$ AU, farther than it did in
simulation Sim1.  This can be understood from equation \ref{dadt2e}. The two
terms on the right hand side of that equation have different signs, with the
last term (due to radiative cooling) causing inward migration. This term
remains roughly the same in Sim1 and Sim2 as $r_p$ is almost the same in the
two simulations. However, the external disc torque is much larger in
Sim2. Since the net result of the outer disc torque is outward migration in
this case, the planet migrates outward faster in Sim2 than it does in Sim1.

We would like to pause here for longer to clarify the physical interpretation
of the faster radial migration in Sim2 compared with Sim1 better. At first
this result seems to contradict the common sense: ``A more massive outer disc
should provide a larger spin down torque, so the planet should migrate outward
slower or even migrate inward in Sim2 as compared to Sim1''. However, the
equilibrium mass loss conditions imply that the planet must lose mass more
rapidly in Sim2 precisely because the outer disc torque is larger. This means
that the inner disc in Sim2 is more massive than it is in Sim1, and it
produces a larger outward torque on the planet. To be in the equilibrium
situation, the outward and inward disc torques on the planet must work in such
a way as to push it steadily outward at a rate proportional to the outward
disc torque (cf. equation \ref{dadt2c0}). So physically, a larger outer disc
torque causes a stronger response from the planet through the inner disc
feeding and that is why the planet migrates outward even faster.

As the result of this more effective outward migration, the planet does not
actually suffers a nearly catastrophic disruption when it reaches $M\approx 1
M_J$. At the start of the second planet mass loss spike (time $t\approx 400$
yrs in fig. \ref{fig:hist_Sim2}), the planet moves outward quickly. Instead of
suffering a runaway mass loss as in Sim1, a shutdown of the mass loss occurs
at this moment in Sim2. The planet continues to be pushed outward by the
inflated inner disc, reaching $R\approx 0.3$ AU at $t\approx 500$ yrs.  By
that time the inner disc drains sufficiently onto the star, and the inward
migration of the planet resumes. Despite some contraction of the planet's
radius with time, the Hills radius becomes smaller than $r_p$ again at time
$t\approx 800$ yrs. This time Roche lobe overflow is terminal for the gaseous
part of the planet, and it is completely destroyed in the third Roche lobe
overflow event very quickly. The evolution of the disc after that is similar
to Sim1.

\subsection{Thermal instability in the gap (Sim3)}\label{sec:sim3}

To explore the importance of initial conditions, we set up simulation Sim3,
which has same initial disc mass as Sim2, $M_d = 50 M_J$, but higher disc
viscosity parameter, $\alpha = 0.01$, and a different disc radial cutoff
$R_0=7$ AU (see equation \ref{sigma0}). The outer boundary of the disc is
enlarged to $R_{\rm out} = 40$ AU. We also place the planet at larger starting
separation, $a_0 = 1$ AU rather than $0.13$ AU in Sim1, Sim2.

\begin{figure}
\centerline{\psfig{file=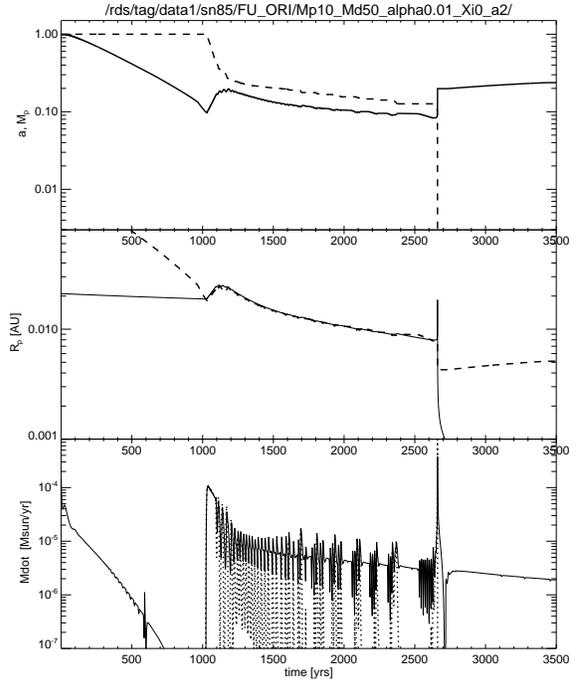,width=0.5\textwidth,angle=0}}
\caption{Same as Fig. \ref{fig:hist_Sim2} but for simulation Sim3, in which
  the planet is inserted at $a = 1$ AU instead of $a=0.13$ AU. The
  oscillations in the mass loss/gain rates in the bottom panel are due to
  thermal ionisation instability in the gap, as explained in \S \ref{sec:sim3}.}
\label{fig:hist_sim3}
\end{figure}

The results of this test are shown in Figure \ref{fig:hist_sim3}.  If we were
to average out the shorter period (tens of years) variations in the curves,
the gross evolution of the system is not that different from that found
previously. Once again, the inner disc drains onto the star whilst the outer
disc is ``dammed up'' behind the planet. The accretion rate onto the star then
plummets to very small values until the planet is pushed to $R\sim 0.11$ AU
where it overfills its Roche lobe. The material siphoned from the planet feeds
the inner disc, restarting the accretion onto the star. Neglecting the
oscillatory pattern for now, the planet loses mass and moves first outward
and then inward until $M_p\approx 1.5 M_J$, all the time satisfying
$r_p\approx r_H$. When the gap is partially filled, there is a very large
outburst that unbinds the residual gaseous mass of the planet. After this, the
gap in the disc shuts completely, and the planet is of no importance for the
disc.

\begin{figure}
\centerline{\psfig{file=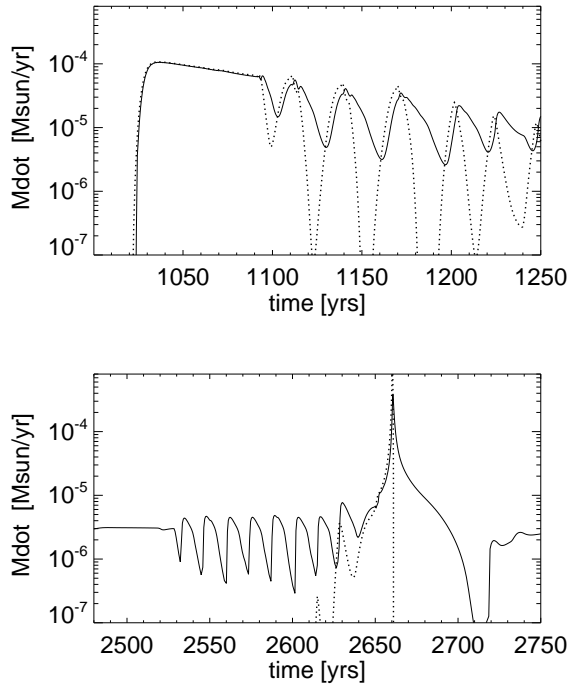,width=0.5\textwidth,angle=0}}
\caption{Same as the bottom panel of Fig. \ref{fig:hist_sim3} but zooming
  onto two interesting time periods in the disc/planet evolution. In the top
  panel, the beginning of the outburst is shown, when the Roche lobe is
  overfilled for the first time. The bottom panel of the figure corresponds to
  the period around the ``final'' destruction of the gaseous envelope of the
  planet.}
\label{fig:hist_sim3_cycle}
\end{figure}

However, on shorter time scales, there is far more variability in Sim3 than in
any presented before. Figure \ref{fig:hist_sim3_cycle} zooms in onto two
selected time intervals from simulation Sim3, showing just the mass loss and
accretion rates. An oscillatory behaviour is seen already $\approx 70$ years
after the the first Roche lobe overflow (cf. the top panel of
fig. \ref{fig:hist_sim3_cycle}). Limit-cycle variations persist throughout
most of the time preceding the ``final destruction spike'' (cf. the bottom
panel of fig. \ref{fig:hist_sim3_cycle}).

\begin{figure}
\centerline{\psfig{file= 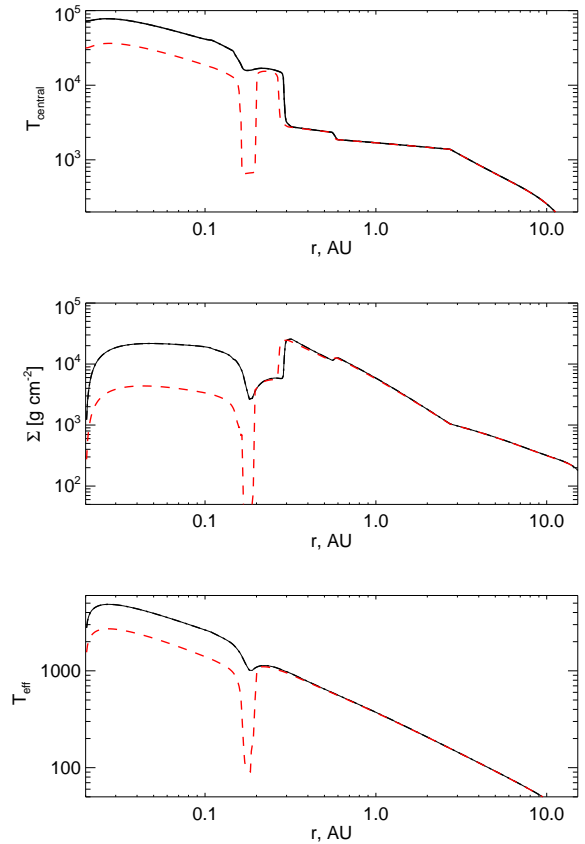,width=0.5\textwidth,angle=0}}
\caption{Disc behaviour through one of the instability cycles for simulation
  Sim3. Top: disc central temperature; Middle: disc surface density; Bottom:
  disc effective temperature. The curves are for two selected times, $t=1171$
  yrs (solid black curves), and $t=1195$ yrs (dashed red curves). These
  correspond to a peak and a trough in the star's accretion rate (cf. \S
  \ref{sec:sim3}).}
\label{fig:temp_sim3_cycle}
\end{figure}

Figure \ref{fig:temp_sim3_cycle} shows the disc central (midplane)
temperature, surface density, and the effective temperature, $T_{\rm eff}$, at
two times, $t=1171$ yrs, and $t=1195$ yrs. The first of these times (solid
black curves in Figure \ref{fig:temp_sim3_cycle}) is close to the third peak
in the stellar accretion rate curve in the top panel of
fig. \ref{fig:hist_sim3_cycle}, whereas the second set of curves (red dashed
curves in fig. \ref{fig:temp_sim3_cycle}, respectively) corresponds to the
next trough in the accretion rate.

Concentrating on the top panel of fig. \ref{fig:temp_sim3_cycle}, we note that
the two curves differ by a factor of $\sim$ two in the inner disc, and are
almost identical behind the gap region. However, the gap region is
significantly different at these two selected times. In particular, the disc
central temperature $T_c \sim 2\times 10^4$ K at the peak of the accretion
rate cycle, and only $T_c \sim 800$ K at the minimum of the cycle.

These profound changes in the disc temperature have similarly profound
consequences for planet evolution and the inner disc, as that is strongly
linked to the planet's migration and mass loss rate. The middle panel
of fig. \ref{fig:temp_sim3_cycle} shows that while the gap is only partially opened
at the ``high-T'' state of the gap region, it gets fully opened at the
``low-T'' state, when viscosity of the material in the gap drops in response
to the temperature variation. 

However, what causes the temperature drop in the gap region? To address that,
note that the planet mass loss rate actually dips much faster than the stellar
accretion rate after any one of the peaks visible in the top panel of
\ref{fig:hist_sim3_cycle}. This implies that the planet has been pushed
outward ``too far'' by the inner disc torques, and it continues to be pushed
even further out for some $\sim 10$ yrs after the peak. The inner disc
eventually runs out of material, as it is accreted onto the star. The outer
disc then over-powers the torques from the inner one and the planet migrates
inward again, restarting the mass loss and re-filling the inner disc for
another cycle.  Thus, while the gap changes are pronounced, they appear to be
driven by the behaviour of the inner disc, which makes the planet to switch
between the full-gap to a partial-opened gap states.

\section{Planet-Disc thermal instabilities}\label{sec:lc04}

During the outburst state, the inner disc becomes very hot, with most of
hydrogen atoms ionised. One may therefore expect that the $\alpha$-parameter
may be higher, in accord with estimates from dwarf nova systems
\cite{KingEtal07}, and Magneto-Rotational Instability simulations
\citep{BH98}. Furthermore, modeling of the observations of FU Ori outbursts
by \cite{ZhuEtal07} suggested that $\alpha \sim 0.02 - 0.2$. We therefore
present now simulation Sim4, which is identical to Sim3 except $\alpha=0.04$.

\begin{figure}
  \centerline{\psfig{file=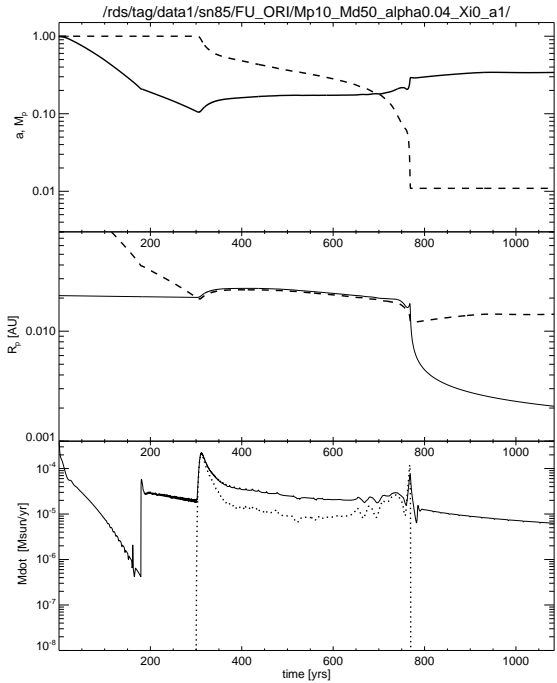,width=0.5\textwidth,angle=0}}
  \caption{Same as Fig. \ref{fig:hist_sim4} but for simulation Sim4, which is
    identical to Sim3 except that $\alpha$-parameter is increased to
    $\alpha=0.04$. The resulting behaviour change is significant: the thermal
    disc instability is triggered before the planet overflows its Roche
    lobe. The resulting disc/planet behaviour is somewhat similar to that found
    by Lodato \& Clarke (2004).}
  \label{fig:hist_sim4}
\end{figure}

Figure \ref{fig:hist_sim4} shows the outcome of this simulation, which
contains both similarities and also very significant differences from all of
the simulations presented earlier. Similarly to the previously studied cases,
the inner disc first empties onto the star, while the outer disc pushes the
planet in. However, unlike any of the previous cases, the first accretion
outburst episode onto the star is not driven by the planet filling its Roche
lobe and transferring its mass inward. Instead, at time $t\approx 179$ yrs,
thermal disc instability is triggered. At that time, the planet is situated at
$R\approx 0.2$ AU and does not fill its Roche lobe. 

\begin{figure}
  \centerline{\psfig{file=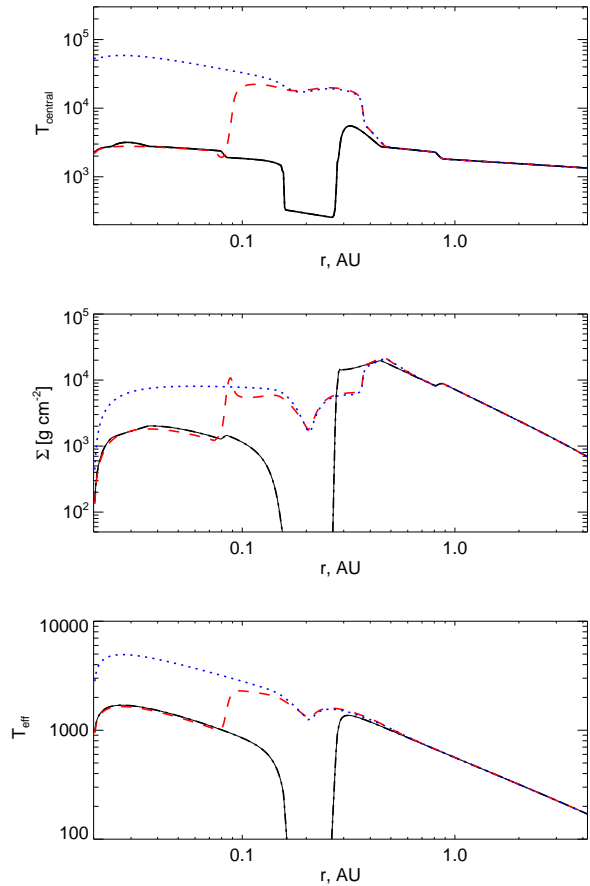,width=0.5\textwidth,angle=0}}
  \caption{The disc structure at times $t=176.4$ yrs, 179.2 and 179.5
    yrs for simulation Sim4, shown with the black solid, red dashed and
    blue dotted curves, respectively. Note that thermal hydrogen
    ionisation instability is triggered just behind the gap, and the
    instability then propagates in, as in models of
    \citep{LodatoClarke04}.}
  \label{fig:temp_sigma_sim4}
\end{figure}

To analyse the disc behaviour in greater detail, Figure
\ref{fig:temp_sigma_sim4} shows the disc structure just before the instability
is triggered (black solid curve), during the initial luminosity rise (red
dashed curve) and just after the rise (blue dotted). One notices that the
instability is actually triggered at the outer banked-up edge of the gap,
behind the planet, at $R\sim 0.3$ AU. Before the instability, the disc is
relatively cold everywhere, but especially so in the gap. As the instability
flares up, the disc behind the gap heats up and the increased pressure of the
gas is able to partially close the gap. The gas thus rushes inward,
accumulating in the inner disc, sending that region of the disc on the hotter,
upper stable branch of the ``S-curve'' \citep{Bell94,LodatoClarke04}, and
leading to a massive rapid-rise outburst.

The outburst in the simulation Sim4 is in a qualitative agreement with
\cite{LodatoClarke04}, except here it is supplemented by the mass loss from
the planet. Note that in this high $\alpha$ (and thus high accretion rate)
simulation, the mass loss from the planet is a relatively minor
detail. Indeed, when the mass loss from the planet sets in ($t\approx 300$
yrs), it exceeds the prior value of the accretion rate onto the star only in
the first $\sim 50$ yrs or so. Together with Sim3, this simulation clearly
demonstrates that non-linear coupling between the hydrogen ionisation
instability in the disc and the planet-disc mass exchange may be important in
determining a variability pattern in the accretion rate onto the star.

\section{Emission region size}\label{sec:size}

\begin{figure}
  \centerline{\psfig{file= 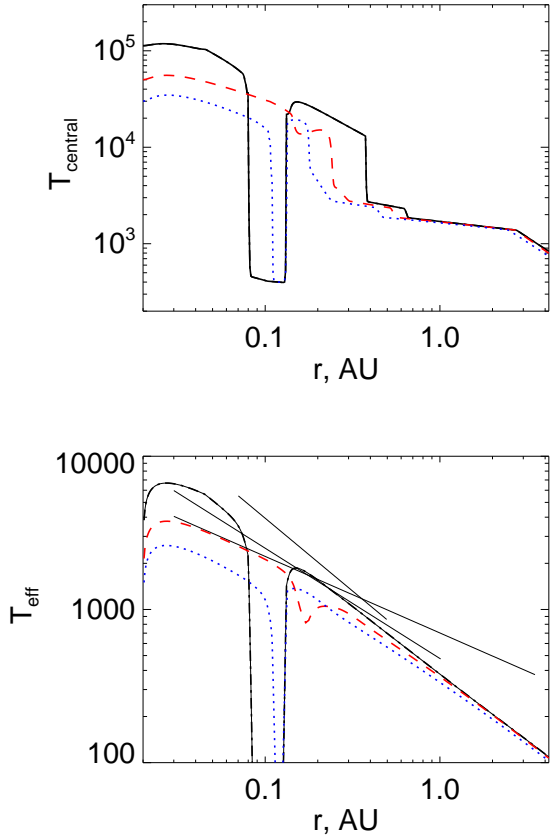,width=0.5\textwidth,angle=0}}
  \caption{The disc central temperature (Top panel) and the effective
    temperature (Bottom panel) versus radius for simulation Sim4, at three
    selected times, $t=1036, 1248$ and 1584 yrs, shown with the thick solid
    black, red dashed and blue dotted curves, respectively. The three black
    solid power-laws represent the best fitting effective temperature profiles
    inferred by Eisner and Hillenbrand (2011) for the three FU Ori sources.}
  \label{fig:temp_sim4}
\end{figure}

Figure \ref{fig:temp_sim4} shows the disc central temperature (top panel) and
the disc effective temperature (bottom panel) for several representative times
for the simulation Sim4. Along with those curves, we also show the best fit
power-laws of the inner hot disc model of \cite{EisnerH11} to
their NIR Keck interferometer data for the three well known FU Ori
objects. These observations have spatial resolution as good as $\sim 0.05$ AU, and
thus serve as direct and sensitive probes of the spectral disc model fits.

The effective temperature profiles of our discs in the inner region are not
that dissimilar from the best fit models of \cite{EisnerH11}, except for the
presence of the gap.  This qualitative agreement is encouraging, given that we
did not try to fine tune our models to satisfy the spectral constraints. This
suggests that our models may be relevant to the observed FU Ori sources,
although it is desirable to attempt more detailed spectral fits to the
observations in the spirit of \cite{EisnerH11}. This is however beyond the
scope of the present paper.

\section{Discussion}\label{sec:disc}

\subsection{Main results of this paper}\label{sec:mr}

In this paper we presented calculations of the tidal disruption of massive
gaseous planets in massive accretion discs of young protostars. In the
analytical part of the paper (\S \ref{sec:analytical}), we derived
conditions under which the mass and angular momentum exchange between
the planet and the disc is in a quasi steady state (cf. text around
equation \ref{unstable}), and the corresponding equilibrium mass loss
rate from the planet. Qualitatively, these conditions require (a) the
planet to be massive enough to keep the gap in the disc opened; and
(b) the mass-radius relation for the planet to be such that the planet
contracts as it loses mass or at least expands only as a weak
power-law of the planet's mass, $M_p$.

We then formulated a numerical 1D approach to modeling the planet-disc
mass and angular momentum exchange (\S \ref{sec:numerics}). Utilising
a toy cooling model for a young planet, and a fixed external torque
driving the planet in, we found that our numerical model reproduces
the analytically expected results well (\S \ref{sec:CD}), including
the quasi-steady state planet's disruption (Roche lobe overflow), and
the eventual closing of the gap and a runaway planet disruption when
the planet's mass becomes  too low. We then proceeded to
several tests (labelled Sim1 to Sim4 in Table 1) in which the
planet-disc torques are calculated self-consistently.

The most robust result of our calculations is that if there is a
massive young planet in the inner disc, then the accretion rate onto
the protostar must be strongly variable on time scales ranging from a
few years to a few hundred or even thousands of years. There are
simply too many non linear physical processes occurring in the
planet-disc system for the accretion rate to remain static over long
time scales.

First of all, as well known from previous literature
\citep{RiceEtal03} a massive planet opens a deep gap in the disc,
blocking the flow of matter from the exterior disc to the inner
one. Such a blockade of the protostar by the young planet may be the
reason why the average accretion rates of protostars appear to be
lower than expected theoretically \citep{DunhamEtal10}.

We found two ways in which the blockade is ended, and both lead to a
rapid rise accretion outburst that may be relevant for the
observations of FU Ori outbursts. The first of these, encountered in
the relatively high $\alpha$ simulation Sim4 ($\alpha=0.04$, cf. \S
\ref{sec:lc04} and figure \ref{fig:hist_sim4}), is the same as the
\cite{LodatoClarke04} model for FU Ori outbursts. In this case thermal
(hydrogen ionisation) instability sets in at the outer edge of the gap
in the material piled up behind the planet. Heating up, the outer disc
is able to close the gap, overflow the planet and fuel an outside-out
accretion outburst.

The second way to kick-start accretion onto the star was found in all
of our lower $\alpha$ simulations (Sim1 to Sim3 in Table 1). In these
cases the planet is pushed inward sufficiently fast to fill its Roche
lobe before the hydrogen ionisation instability at the outer gap takes
place. The planet then loses mass through its L1 point. The material
circularises inward of the planet, and spreads viscously. As gas
accretes onto the protostar, the accretion rate rises by orders of
magnitude. What happens next depends on the internal properties of the
planet. If conditions for the quasi-steady state mass transfer are
satisfied, then the planet migrates radially (usually in the outward
direction), so that the Hills radius is exactly equal to the planet's
radius. The resulting outburst light curve may then resemble a step
function, similarly to the FU Ori outburst itself \citep{CLEtal07}.

During this quasi-steady state planet-star mass transfer, the role of
the planet completely reverses. While before the outburst the planet
prevented the material from the outer disc from reaching the star, now
it is the donor for the star. Also, like in a stellar binary system,
the planet takes the excess angular momentum of the inner disc away and does
not allow the material to spread outward viscously. This latter effect
keeps the accretion rate onto the star at a higher value for longer
than it would have been if the gas could flow past the planet.

If the steady-state conditions for mass transfer are violated, then
an even more dramatic outburst may occur. The planet is then tidally
disrupted on time scale of a few to few tens of years. The material
disrupted off the planet then spreads viscously in both directions,
and the accretion rate onto the star drops rapidly as there is no
planet to bank up the inner disc in the outward direction.

On the top of this generic behaviour, we found variability on smaller
time scales caused by a non-linear connection between the planet
migration and mass loss and hydrogen ionisation instability. For
example, in simulation Sim3 (\S \ref{sec:sim3}), only the gap region
of the disc is unstable and switches between the ionised and the
non-ionised states, modulating the accretion rate onto the protostar.

These results do depend quite sensitively on the internal structure of
the planet, as we described in \S \ref{sec:embryo_contraction}. In
particular, if planets are smaller after H$_2$ dissociation than we
assumed in most of our simulations here, then they may not be
disrupted at all. In this case the planets are driven all the way into
the star or the magnetospheric cavity. This corresponds to the ``high
density start'' models \citep{Nayakshin11b} parametrised by $X_i =
1$. Therefore, better modeling of the internal relaxation of the
growing planets, including possible massive solid cores and the
luminosity released by these, is absolutely necessary to detail
predictions of our model further.

\subsection{Where do the inner planets come from?}\label{sec:new} 

It is possible that every young protostar goes through as many as
10-20 FU Ori events during its growth \citep{HK96}. Recent modeling
of the embedded phase of low mass proto stars by \cite{DunhamEtal10}
also suggests that large amplitude variability in the accretion rate is
widespread, and that stars may actually acquire most of its mass in
bursts. If our model for FU Ori outbursts is to apply to the
observations, many giant planets are required per star. Is this
reasonable?

The answer strongly depends on which planet formation theory is accepted. In
the standard Core Accretion (CA) model \citep{PollackEtal96,AlibertEtal05} of
planet formation, giant planets are formed ``late'', e.g., after a few
Myrs. These planets are also expected to be rather compact
\citep{MarleyEtal07}. Tidal disruption of giant CA planets is therefore
unlikely, and having as many as ten of such per star would also seem
implausible.

The gravitational disc instability model for planet formation
\citep[e.g.,][]{Boss97} was thought to form planets only at distances greater
than 50-100 AU \citep[e.g.,][]{Rafikov05} due to inefficient cooling at closer
distances \citep{Gammie01,MayerEtal04,Rafikov05,Meru10}. It may therefore
appear that planets born at these large distances would be of no relevance to
this paper. However, working in the context of proto stellar accretion rather
than planet formation, \cite{VB05,VB06} showed that their clumps migrated
inward rapidly, reaching all the way to their inner computational domain
boundary of 10 AU. These results have been supported by their more recent
simulations \citep{VB10} and by simulations of a number of independent authors
\citep{BoleyEtal10,BoleyDurisen10,ChaNayakshin11a,MachidaEtal11,BaruteauEtal11,MichaelEtal11}. \cite{BaruteauEtal11}
provide an in depth analysis of the rapid migration of gas clumps in
self-gravitating discs and conclude that these clumps migrate via type I
regime. These authors find that massive planets migrate inward faster than a
disc gap could be opened. A recently submitted manuscript by Zhu et al (2011b;
private communication) also shows many examples of massive clumps migrating
inward rapidly although the authors find that some of these clumps become
substantially more massive than a giant planet.

Note that these results only apply to the outer gravito-turbulent discs;
  in the inner non self-gravitating disc regions studied here, planet
  migration must slow down considerably just because planets we consider are
  1-3 orders of magnitude more massive than the total mass of these inner
  discs. Standard gap opening criteria should apply then.

\cite{BoleyEtal10,Nayakshin10c} argued that tidal disruption of
gaseous envelopes of the giant proto-planets may be a plausible way to
form all kinds of planets in what was called the ``Tidal Downsizing
Hypothesis'' for planet formation. More recently, \cite{Nayakshin11b}
have shown that proto-planets in which molecular hydrogen is
dissociated are dense enough to undergo tidal disruption inside the
inner $\sim 0.1$ AU from the star.

If these ideas are to be relevant to FU Ori outbursts and episodic accretion
of stars, one must accept that most early formed planets are completely
destroyed and swallowed by their parent stars. This is required both from
the fact that many outbursts are needed per star \citep[][]{HK96}.

\begin{table*}
\begin{center}
\begin{tabular}{|c|c|c|c|c|c|c|c|c|c|c|c|}
\hline \hline
Name$^{\, a}$ & $M_d$ $^{\, b}$ & $\alpha^{\, c}$ & $X_i$ $^{\, d}$ & 
 $a_0^{\, e}$ & $\zeta_p^{\; f}$ & & Figs.$^{\, g}$ & Comments$^h$\\
& $\left(M_J\right)$ &  & & $\left(\hbox{AU}\right)$ &  &  & & 
\\ \hline \hline

Ext1 & -- & 0.02 & 0 & 0.25 &  $-1/3$ & & \ref{fig:CD_hist},\ref{fig:CD_disc}
& Reference case\\ \hline
Ext2 & -- & 0.1 & 0 & 0.25 &  $-1/3$ & & \ref{fig:CD_hist2} &
Gap closed\\ \hline
Ext2 & -- & 0.02 & 0 & 0.25 &  $-1$ & & \ref{fig:Ext3} &
Unstable mass transfer\\ \hline
 \hline
Sim1 & 20 & 0.002 & 0 & 0.13 &  $-1/3$ & &
\ref{fig:hist_Sim1},\ref{fig:S1_disc1}, \ref{fig:histZoom_Sim1}
& Reference case\\ \hline
Sim1.X & 20 & 0.002 & X & 0.13 &  $-1/3$ & &
\ref{fig:hist_Xi0.4},\ref{fig:hist_compXi}
& closer-in disruptions\\ \hline
Sim2 & 50 & 0.002 & 0 & 0.13 &  $-1/3$ & &
\ref{fig:hist_Sim2}
& Temporary shutdown of mass transfer\\ \hline
Sim3 & 50 & 0.01 & 0 & 1 &  $-1/3$ & &
\ref{fig:hist_sim3}, \ref{fig:hist_sim3_cycle}, \ref{fig:temp_sim3_cycle}
& Thermal instability in the gap \\ \hline
Sim4 & 50 & 0.04 & 0 & 1 &  $-1/3$ & &
\ref{fig:hist_sim4}
& Disc instability before mass loss \\ \hline
\end{tabular}
\end{center}
\caption{List of simulations performed and main parameters. Notes:}

\begin{flushleft}
$^a$ Simulation IDs. Those starting with ``Ext'' are simulations with an
  imposed external torque (\S \ref{sec:CD}). Those with ``Sim'' are full disc
  simulations where the disc-planet torques are calculated self-consistently.\\
$^b$ Initial disc mass for the full disc simulations. \\
$^c$ $\alpha$-parameter of the simulation.\\
$^d$ The fraction of ionised hydrogen used to determine the initial planet
  radius (eq. \ref{re0_est}).\\
$^e$ The starting location of the planet, in AU.\\
$^f$ The index of the planet's mass-radius relation (equation \ref{zetap}).\\
$^g$ Corresponding figures in the paper.\\
$^h$ Notable behaviour in the simulation.\\

\hrulefill
\end{flushleft}

\label{table_results}
\end{table*}

\section{Conclusions}\label{sec:conclusions}

In this paper we studied the planet and the disc evolution in cases when the
young massive gas giant planet overfills its Roche lobe and transfer its mass
back into the disc. Due to significant uncertainties in the internal structure
of young planets, large parameter space for the problem, and still missing
physics, our study cannot yet give definitive answers on what exactly happens
with the planets. Instead, this work is to be viewed as one establishing an
analytical and numerical framework for further studies of the planet-disc-star
mass exchange.

It appears that Roche lobe overflow of young giant gas planets holds a
potential promise as an explanation for the enigmatic FU Ori outbursts of
protostars. We note in passing that a subset of our models (such as Sim3) may
be relevant to shorter period and lower accretion rate variable protostars
such as EXors \citep{Herbig89,SAEtal08,LorenzettiEtal09}. If FU Ori outbursts
and episodic accretion onto protostars are indeed connected to tidal
disruptions of young planets then the implication would be that many young
giant planets are tidally destroyed or perhaps swallowed whole by their parent
proto-stars.

Future work should include better treatment of the planet's internal
  structure, especially a possible presence of truly massive solid cores, add
  type I migration (neglected here), the possibility for the overflow through
  the L2 point (neglected here), and aim to model the ``end'' result when the
  gas disc is presumably dispersed by disc photo-evaporation.

\section{Acknowledgments}

Theoretical astrophysics research in Leicester is supported by an STFC Rolling
Grant. This research used the ALICE High Performance Computing Facility at the
University of Leicester.  Some resources on ALICE form part of the DiRAC
Facility jointly funded by STFC and the Large Facilities Capital Fund of
BIS. Cathie Clarke's careful reading of the manuscript and useful suggestions
are much appreciated. Andrew King is thanked for discussions of mass transfer
in stellar binaries. Richard Alexander is thanked for useful comments on the
early draft of the manuscript. Charles Gammie is thanked for showing us the
Zhu et al. 2011b manuscript before it appears in the public domain. SN thanks
Rashid Sunyaev and the Max Planck Institute for Astrophysics in Garching where
the manuscript was completed for hospitality. Finally, we benefited from
discussions of planet migration with Sijme-Jan Paardekooper and Clement
Baruteau.


\label{lastpage}

\end{document}